\documentclass[12pt, a4paper]{article} 
\usepackage[utf8]{inputenc}
\usepackage{dcolumn,lscape}
\usepackage{graphicx}
\usepackage{natbib} 
\usepackage{url} 
\usepackage[table]{xcolor} 
\usepackage{amsmath,amssymb}
\usepackage{hhline}
\usepackage{graphicx}
\usepackage{arydshln}
\usepackage[format=hang,labelfont=bf,textfont=small,singlelinecheck=false,justification=raggedright,margin={12pt,12pt},figurename=Fig.]{caption}
\usepackage{titlesec}
\usepackage{textcomp}
\usepackage{amsmath,amssymb,dsfont, mathtools}
\usepackage{makecell}
\usepackage{multirow}
\usepackage{exscale,longtable,multicol,dcolumn,tabularx,bm}
\usepackage{adjustbox}
\usepackage{neuralnetwork}
\usepackage{amsmath,longtable,multicol,dcolumn,tabularx,graphicx,amssymb}
\usepackage{exscale,amsthm,multirow,rotating}
\usepackage{bbm}
\usepackage{array}
\usepackage{comment}
\usepackage{tikz,dsfont,float}
\usepackage[utf8]{inputenc}
\usepackage{hyperref}
\hypersetup{
    colorlinks=true,
    citecolor = black,
    linkcolor=blue,
    filecolor=magenta,      
    urlcolor= blue 
}
\usepackage{bbm}
\usepackage{tikz,dsfont,float}
\usepackage{mathptmx} 
\usepackage{mathrsfs}
\usepackage{booktabs} 
\usepackage{subcaption}

\usepackage{lscape}

\DeclareMathOperator*{\argmax}{arg\,max}

\usetikzlibrary{calc}
\usetikzlibrary{positioning}

\usepackage{graphicx}
\usepackage{amsmath,amssymb,dsfont, mathtools}
\usepackage{natbib}
\usepackage{makecell}
\usepackage{multirow}
\usepackage{exscale,longtable,multicol,dcolumn,tabularx,bm}
\usepackage{adjustbox}
\usepackage{booktabs} 
\usepackage{mathptmx}      
\usepackage{mathrsfs}

\usepackage{xcolor, ulem}
\begin{document}

\def\spacingset#1{\renewcommand{\baselinestretch}%
{#1}\small\normalsize} \spacingset{1}


  \title{\bf Statistical process monitoring of artificial neural networks}
  \author{Anna Malinovskaya\\
  	\small{Leibniz University Hannover, Institute of Cartography and Geoinformatics, Germany}\\
  	Pavlo Mozharovskyi\\
  	\small{LTCI, Télécom Paris, Institut Polytechnique de Paris, France}\\
  	Philipp Otto\\
  	\small{Leibniz University Hannover, Institute of Cartography and Geoinformatics, Germany}}
  \maketitle
\begin{abstract}
The rapid advancement of models based on artificial intelligence demands innovative monitoring techniques which can operate in real time with low computational costs. In machine learning, especially if we consider artificial neural networks (ANNs), the models are often trained in a supervised manner. Consequently, the learned relationship between the input and the output must remain valid during the model's deployment. If this stationarity assumption holds, we can conclude that the ANN provides accurate predictions. Otherwise, the retraining or rebuilding of the model is required. We propose considering the latent feature representation of the data (called ``embedding'') generated by the ANN to determine the time when the data stream starts being nonstationary. In particular, we monitor embeddings by applying multivariate control charts based on the data depth calculation and normalized ranks. The performance of the introduced method is compared with benchmark approaches for various ANN architectures and different underlying data formats. 
\end{abstract}

\noindent%
{\it Keywords:} Change Point Detection, Data Depth, Latent Feature Representation, Multivariate Statistical Process Monitoring, Artificial Neural Networks, Online Process Monitoring.
\vfill

\spacingset{1.45} 

\section{Introduction}
\label{sec:intro}

The rapid advancement of Artificial Intelligence (AI) has encouraged practitioners in various scientific fields to examine the benefits and challenges of Machine Learning (ML) algorithms in their research. For instance, in astronomy, the separation of astrophysical objects such as stars and galaxies can be performed by applying decision trees \citep{ball2006robust}; reliable forecasts of energy consumption using support vector machines are extensively studied in civil engineering \citep{gao2019machine}; in biology, artificial neural networks (ANNs) are applied for predicting molecular traits \citep{angermueller2016deep}. Due to the emergence of big data and the increased capability of computers, the development of ANNs has received particularly broad attention (cf. \citealp{chiroma2018progress}). Current ANNs, consisting of many hidden layers and being called ``deep learning'' models, have achieved impressive results in many areas, such as finance, linguistics, or photogrammetry (cf. \citealp{aldridge2019neural, otter2020survey, heipke2020deep}). 

Considering supervised learning for tasks such as regression or classification, data with a known relationship between the input and the output is required. During training, the algorithm minimizes the error between the model's output and the target. After the parameter estimation and the testing phase, the model is deployed on new data, promising a stable performance. However, if the distribution of input data changes, it would violate the stationarity assumption and adversely impact predictive performance (see also \citealp{huang2011adversarial}).

In computer science, ``concept drift'' refers to the problem of changes in the data distribution that render the current prediction model inaccurate or obsolete \citep{demvsar2018detecting}. In statistics, ``nonstationarity'' describes time series with changing statistical properties. Additionally, ``novelty'' can occur, e.g., when the data stream introduces instances of a new class, being not present during training (cf. \citealp{masud2009integrating,garcia2019online}). Model revision is necessary to address these issues, ranging from retraining with new data to algorithm replacement. Strategies handling nonstationarity without explicit detection exist (cf. \citealp{gama2014survey}), but current research highlights the benefits of employing drift detection methods during model deployment (cf. \citealp{piano2022detecting, zhang2023concept}). This minimizes redundant adjustments while maintaining prediction accuracy.

From a statistical point of view, for real-time detection of changes, we can apply online monitoring techniques, {e.g.}, univariate or multivariate control charts (cf. \citealp{celano2013performance, psarakis2015adaptive, perdikis2019survey}). These monitoring methods are primary techniques in Statistical Process Monitoring (SPM), testing for unusual variability over time \citep{montgomery2020introduction}. The combination of control charts and ML algorithms has become a prominent concept for improving SPM procedures (cf. \citealp{psarakis2011use, weese2016statistical, lee2019process, apsemidis2020review, zan2020statistical, sergin2021toward, yeganeh2022ensemble}). However, the authors are not aware of competitively using control charts in reverse to supervise the quality of ANN applications in a realistic setting, e.g., considering the limited availability of labels during model deployment and various ANN architectures. Consequently, statistical monitoring of ANN applications is a field offering new perspectives for SPM.

In an ANN, data flows through hidden layers, each containing neurons that perform nonlinear transformations, ultimately producing predictions in the output layer \citep{hermans2013training}. The hidden layers combine learned data representations from previous layers, abstracting irrelevant details and extracting useful features for generating the output. Neurons within the ANN, as demonstrated by \cite{wang2019explaining}, possess two crucial properties: stability and distinctiveness. For smooth classification surfaces or prediction functions, nearby samples should activate similar neurons, leading to similar output values. However, in the presence of nonstationary data, such as from an unknown class, neuron values may deviate beyond the typical activation range. In this case, its interim representation generated by ANN before the output layer would be different compared to the interim representation of the data that correctly belongs to a predicted class. Commonly represented by a low-dimensional vector, the learned latent feature representation also called ``embedding'' comprises a dense summary of the incoming sample. Alternatively, sliced-inverse regression can reduce dimensionality without specifying a regression model \citep{li1991sliced}. It projects predictors onto a subspace while preserving information about the conditional distribution of the response variable(s) given the predictors.
Various methods have been proposed that utilize the inverse relationship between the variables (\citealp{cook2005sufficient, wang2008sliced, wu2008kernel}), or sparse techniques facilitating the interpretation (\citealp{li2006sparse, li2007sparse, lin2019sparse}). Both ways of reducing the dimensionality and compressing the knowledge about a data sample are suitable for SPM.

The majority of multivariate SPM methods are based on the assumption that the observed process follows a specific distribution. In the general case of ANNs, however, the distribution of embeddings is unknown. Although the network could be trained in a way such that the embeddings follow a certain distribution, this concept is too restrictive in practice. In this work, we focus on developing a nonparametric SPM approach using a data depth-based control chart, making no assumptions about the ANN type and the distribution of embeddings.

 In Section \ref{sec:background}, we review the relevant literature and describe the notation related to ANNs, followed by the definition of the studied change point detection problem. Afterward, we discuss the notion of data depth and respective control charts in Section \ref{ProposedMethod}. The experimental results of our approach and comparison to the benchmark methods on both synthetic and real data are provided in Section \ref{sec:comparativestudy}, followed by the discussion about open research directions in Section \ref{sec:conclusion}.

\section{Background and Notation}
\label{sec:background}

Detecting nonstationarity can be explored from different perspectives. In statistics, we would usually look at change point detection methods (cf. \citealp{ali2016overview}), while in computer science, these techniques are rather known as concept drift, anomaly, or out-of-distribution detection (cf.  \citealp{vzliobaite2016overview, yang2021generalized, fang2022out}). In the following sections, we briefly review the relevant literature from different fields (Section \ref{LitReview}) and introduce important notation of ANNs (Section \ref{ANN}) together with the change point detection framework (Section \ref{CPD}).

\subsection{Literature Review} \label{LitReview}
In general, approaches to detect nonstationarity in ML applications can be subdivided into two groups: performance-based methods and distribution-based methods \citep{hu2020no}. The first group relies on labeled instances,  monitoring the classification error rate or other performance metrics (cf. \citealp{klinkenberg1998adaptive, klinkenberg2000detecting, nishida2007detecting}), for example, by conducting sequential hypothesis tests based on the ideas of control charts (cf. \citealp{gama2004learning, baena2006early, kuncheva2009using, mejri2017combination}). \cite{mejri2021new} developed a two-stage time-adjusting control chart for monitoring misclassification rates, which updates the control limits in the first stage, and validates the detected changes in the second stage. However, in practice, labeled data is usually not available when ANNs are applied. Thus, the idea of adapting performance-based approaches to a confidence-related output produced by a model was proposed \citep{haque2016efficient, kim2017efficient}. If a classifier processes the anomalous data, it is expected that the model's confidence about the affinity of a data point to a certain class would change so that the unusual behavior could be detected (cf. \citealp{hendrycks2016baseline}).

Anomaly detection in distribution-based methods involves the analysis of metrics related to the distribution. A considerable number of techniques apply a sliding window approach on either a one-dimensional data stream or on several features individually, comparing the data distribution of the current window to the reference sample obtained from the training dataset (cf. \citealp{bifet2007learning, bifet2018machine, gemaque2020overview}). However, the underlying reason related to the notable performance of ANNs is their generalization ability in complex high-dimensional AI tasks such as object or speech recognition \citep{goodfellow2016deep}, meaning that the detection of nonstationarity from the original data would not be appropriate due to the excessive number of features.  Considering novelty detection, it can be based on parametric density estimates (e.g., using Gaussian mixture models \citep{roberts1994probabilistic} or hidden Markov models \citep{yeung2003host}), and nonparametric estimates (e.g., based on k-nearest neighbors \citep{guttormsson1999elliptical}, kernel density estimates \citep{yeung2002parzen} or string matching \citep{forrest1994self}). These methods usually require heuristically chosen thresholds to decide about the novelty. The interested reader may refer to \cite{markou2003noveltyA, markou2003noveltyB} for a more detailed review.

By considering a latent representation of the original data stream such as embeddings generated by ANNs, outlier and anomaly detection methods based on distance metrics and nearest neighbor approaches were shown to be suitable and efficient in detecting nonstationary samples (cf. \citealp{lee2018simple, sun2022out}). Particularly beneficial is their capability of providing an overall outlying score that would consider all data features together, leading to a more explicit decision about the observed abnormalities.

Alternatively, other ML algorithms for drift detection such as Support Vector Machine (SVM) \citep{krawczyk2015one} or autoencoders as specific types of ANNs \citep{pidhorskyi2018generative} can be applied for change detection. Another suggestion is to use ensemble models consisting of, for instance, several decision trees with a majority voting mechanism \citep{li2015learning}. In particular cases, the detection methods can be enhanced through large-scale pre-training techniques, leading to remarkably informative embeddings \citep{fort2021exploring}. However, in our work, we relax any assumptions on the availability of additional data or specific model architectures. Thus, we describe ANNs from a general perspective in the next section.

\subsection{Artificial Neural Networks} \label{ANN}
The main goal of training ANN is to estimate the parameters $\vartheta$ of a highly nonlinear function $f(\cdot, \vartheta): \mathbb{R}^{d} \rightarrow \mathbb{R}^{v}$, mapping a $d$-dimensional input (i.e., observed data) to a $v$-dimensional output (i.e., class labels). The proposed monitoring technique can also be applied to ANNs, which solve regression tasks by grouping the predicted values into a set of classes.  In both cases, the dependent variables are class labels $y_t \in \{1, \dots, v\}$, where the $v$-dimensional output of the algorithm contains the discriminant scores for each of the given classes. In supervised learning, we assume that true class labels are known for the training period $t = 1, \ldots, T$, which is used for estimating the parameters $\vartheta$. Thus, $\hat{y}_t = \argmax f(\bm{x}_t, \hat{\vartheta})$ for a set of input variables $\bm{x}_t$ is the prediction of ANNs, i.e., the most probable class, where $\hat{\vartheta}$ defines the learned parameters during the training, such that $\hat{y}_t$ coincides with $y_t$ in most cases for all $t = 1, \dots, T$.

\begin{figure*}[t!]
     \centering
     \begin{subfigure}[b]{0.5\textwidth}
         \centering
         \includegraphics[width=0.8\textwidth, trim= 0cm 0cm 22.5cm 0cm,clip]{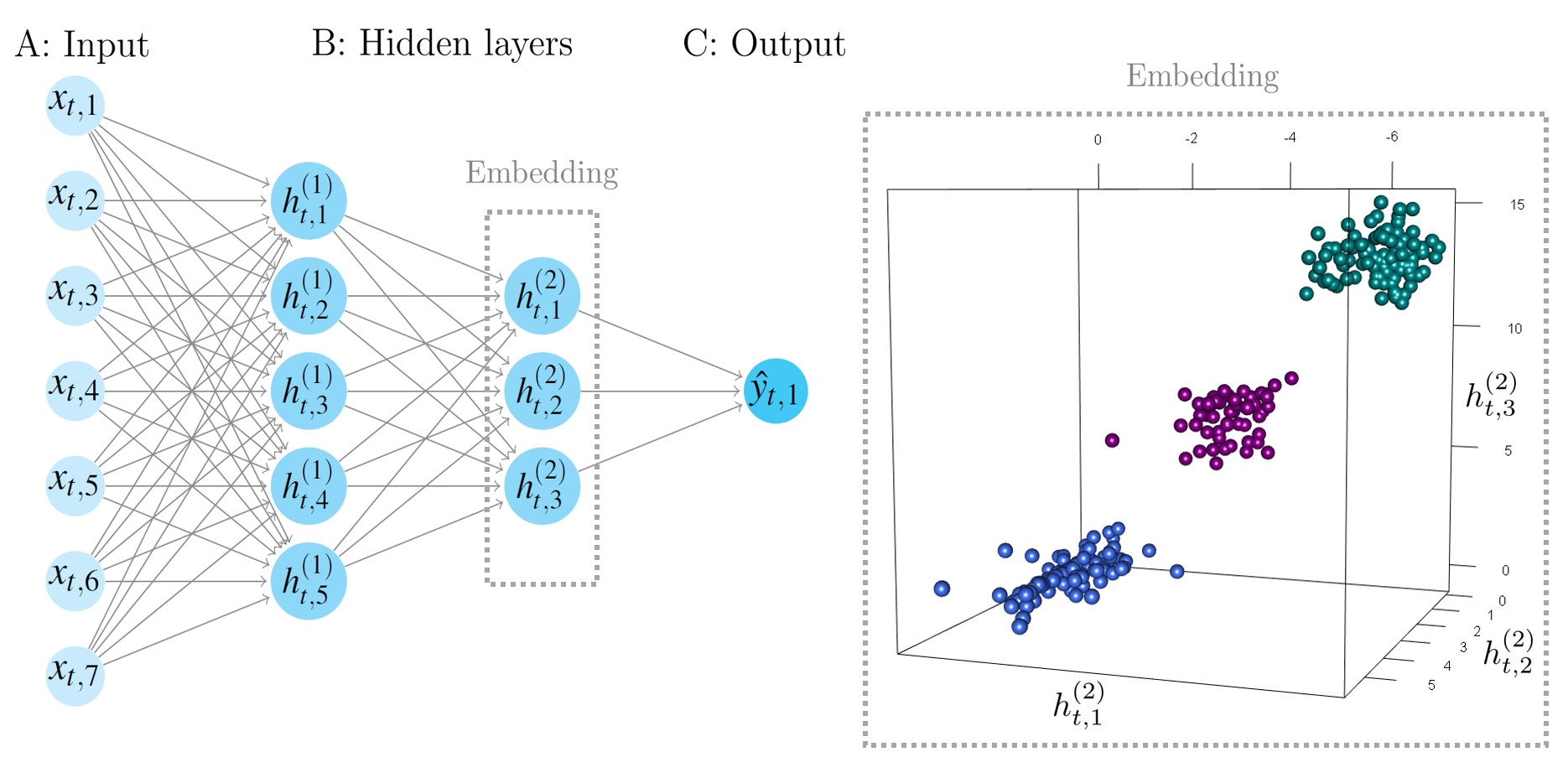}
         \caption{The FNN with two hidden layers in toy example}
         \label{a:FNN}
     \end{subfigure}
     \hfill
     \begin{subfigure}[b]{0.45\textwidth}
         \centering
         \includegraphics[width=0.8\textwidth, trim= 27cm 0cm 0cm 0cm,clip]{CombinedFigures_Fig1.jpg}
         \caption{Produced embeddings in toy example}
         \label{b:vis}
     \end{subfigure}
        \caption{The FNN architecture displayed in (\textbf{a}) and embeddings from hidden layers (B) visualized in (\textbf{b}). Gray-framed nodes represent the last hidden layer. Blue and green points denote training data references, while magenta points represent out-of-control data embeddings.}
        
        \label{NNforward_combined}
\end{figure*}

There are several aspects to consider about the network's architecture when designing it, for instance, the number of hidden layers and neurons, the connections between the layers, and, fundamentally, which type of ANNs to use. To introduce these essential elements, a toy example of a feedforward ANN (FNN) which defines the basic family of ANN models 
is presented in Figure \ref{a:FNN}. The network consists of four layers with two hidden layers, where each circle represents a neuron or node that stores a scalar value. Consequently, a layer is a $k_i$-dimensional vector with $k_i$ being the number of nodes contained in the $i$-th layer. Here, the input layer is a $k_1$-dimensional vector $\bm{x}_t \in \mathbb{R}^7$ (first layer), and its first neuron is defined as $x_{t,1}$. Assuming we have a classification problem with two classes, we would construct an ANN with a one-dimensional output layer (last layer) that provides a score for the input to belong to class 2. When the output exceeds a threshold (often 0.5), the input is predicted as class 2, otherwise class 1. For a detailed description of the toy example set-up, see Section \ref{ToyExample}.

The changes in the number of nodes, layers, and types of connections lead to new ANN architectures. However, specialized types of ANN models were developed to handle high-dimensional data efficiently. For instance, convolutional ANNs enable image data processing, while recurrent ANNs are suitable for working with temporal sequences (cf. \citealp{shrestha2019review}). It is worth noting that designing ANNs usually follows the trial-and-error principle \citep{emambocus2023survey}. Nevertheless, some strategies are recommended by the experts, e.g., how to start choosing hyperparameter values (cf. \citealp{smith2018disciplined}). To obtain a broader overview of how to design and train ANNs, the reader can refer to \cite{goodfellow2016deep}.

Figure \ref{a:FNN} illustrates how hidden layers in an ANN can reduce the dimensionality of the input sample. That produces valuable interim representations known as ``embeddings'' which we obtain before the activation function is applied, compressing the knowledge about the samples as shown in Figure \ref{b:vis}. They serve as a base for the change point detection procedure explained below.

\subsection{Change Point Detection} \label{CPD}

For monitoring ANN applications, we propose to use embeddings that usually have a vector form which we denote by $\bm{m}_t \in \mathbb{R}^k$ with $k$ being the dimension of the hidden layer that produces the embeddings. This representation is observed every time ANNs are applied, i.e., for historical (training) and new data. Thus, embeddings implicitly depend on $\bm{x}_t$, but also on the fitted parameters $\hat{\vartheta}$, so that changes associated with ANN's data quality and performance can be detected.

In succeeding parts, we refer to the set  $\{\bm{m}_t : t = 1, \ldots, T \}$ as historical data with correctly known labels $\{y_t : t = 1, \ldots, T \}$ and to the set $\{\bm{m}_{i} : i = T+1, T+2, \ldots \}$ as incoming data instance with the predicted class label $\hat{y}_{i}$ obtained from $f(\bm{m}_i, \hat{\vartheta})$. Moreover, we consider that the true label of $y_{i}$ is not available and historical data are stationary. Additionally, we assume that $\bm{m}_t$ follows a certain distribution $F_{\bm{m}_t|y_t = c}$ depending on the true class $y_t$. These conditional distributions $F_{\bm{m}_t|y_t = c}$ are denoted by $\Xi_{c}$ for all classes $c \in \{1, \dots, v\}$. Based on historical data, it is possible to estimate the distribution $\Xi_{c}$ empirically that can be used to determine whether a possible change in the data stream occurred, i.e., whether the current observation $\bm{m}_i$ is generated from a different distribution than $\Xi_{y_{i}}$. Hence, we define a change point $\tau$ to arise if

\begin{equation*}\label{eq:cp_model}
\bm{m}_{i} \quad \sim \quad \left\{ \begin{array}{cc}
\Xi_{y_{i}} &  \text{ if } i < \tau \\
\Xi_{\nu} & \text{ if } i \geq \tau .\\
\end{array} \right. \,
\end{equation*}
In other words, a change point $\tau$ is the timestamp when the analysis of an embedding generated from the incoming data indicates that the data sample belongs to a different unknown distribution $\Xi_\nu$.
Consequently, we can regard this situation as a change in the class definition and formulate a sequential hypothesis test of each sample $i$ for detecting changes in the ANN application as follows
\begin{eqnarray*}
H_{0, i}: && \, \Xi_{y_{i}} = \Xi_{\hat{y}_i}  \qquad \text{against} \\
H_{1, i}: && \, \text{there is a location shift and/or a scale increase in } \Xi_{y_{i}} \,.
\end{eqnarray*}
The alternative hypothesis could be true due to (1) misclassification by the model, i.e., $\hat{y}_i \neq y_{i} = a$, where $a \in \{1,\dots, v\}$, leading to $\Xi_{\hat{y}_i} \neq \Xi_{a}$, or (2) nonstationarity of the data stream, i.e., $\hat{y}_i \neq y_{i} = b$ with $b \notin \{1,\dots, v\}$ and $\Xi_{\hat{y}_i} \neq \Xi_{b}$ because $i \geq \tau$. The accurate distinction between those cases is crucial for reliable change point detection. Labeled data in Phase II helps achieve this, but a practical solution without labeled data is discussed in Section \ref{subsec:misclassif}.

To test repeatedly whether the process is in control over time, i.e., whether the data assigned to a particular class does not deviate from the rest in this class, we can apply a multivariate control chart. Since the class distributions $\{\Xi_c :c = 1,\dots,v\}$ are unknown and can be estimated only empirically, we need a nonparametric monitoring technique that relaxes any distributional assumptions. Alternatively, the nonparametric kernel density estimates can be used, so-called Parzen window estimators (\citealp{parzen1962estimation, breiman1977variable}). Moreover, kernel density estimates also have been proven beneficial for classification tasks (e.g., \citealp{ghosh2006classification}).

Several multivariate nonparametric (distribution-free) charts are based on rank-based approaches. These approaches can be divided into two categories: control charts using longitudinal ranking and those employing cross-component ranking \citep{qiu2014introduction}. The first group includes componentwise longitudinal ranking \citep{boone2012two}, spatial longitudinal ranking \citep{zou2012spatial}, and longitudinal ranking by data depth \citep{liu1993quality}. While the first two subgroups are moment-dependent, control charts based on data depth offer flexibility by not imposing moment requirements. For instance, geometric and combinatorial depth functions, such as Simplicial depth and Halfspace depth, are considered and discussed in Section \ref{depthfunctions}. 

The depth-based control charts allow simultaneous monitoring for location shifts and scale increases in the process. The depth functions discussed in this work are affine invariant and satisfy important axioms such as monotonicity, convexity (except for Simplicial depth), and continuity \citep{mosler2022choosing}. These characteristics make them suitable for our problem, leading us to employ nonparametric control charts based on data depth for online monitoring.

\section{Monitoring Framework for ANN}\label{ProposedMethod}

As discussed in Section~\ref{sec:background}, ANNs input space is usually too complex to identify distributional changes directly from the input data (layer A, Figure \ref{a:FNN}), whereas output does not always contain enough information for this purpose as we demonstrate in Section \ref{sec:comparativestudy} (layer C, Figure \ref{a:FNN}). A sufficient but not excessive amount of information can be obtained by intercepting the input propagation on an intermediate layer of the neural network (layer B, Figure \ref{a:FNN}), e.g., to estimate prediction uncertainty which requires rarely accessible supervised training data (see \citealp{corbiere2019addressing}). While several layers can be considered to account for more complex dependencies, one would normally take those which are closer to the output, achieving the highest dimensionality reduction (see, e.g., \citealp{ParekhMFAB21}). Outputs of the intermediate layers constitute a Euclidean space, where---in the unsupervised setting---anomaly-detection techniques can be applied. These can constitute a neural network themselves (e.g., autoencoder), belong to (statistical) ML like a Local Outlier Factor~\citep{breunig2000lof}, one-class SVM~\citep{ScholkopfPSTSW01}, isolation forest~\citep{liu2008isolation}, or be based on data depth as proposed in this work.

The application of data depth in quality control was originally introduced by \cite{liu1993quality}, resulting in the design of Shewhart-type multivariate nonparametric control charts based on the Simplicial depth \citep{liu1990notion, liu1995control}. According to recent publications on data depth-based control charts (cf. \citealp{cascos2018control,barale2019nonparametric, pandolfo2021multivariate}), the careful choice of data depth is crucial for satisfactory monitoring performance. Thus, we compare several notions of data depth and discuss their effectiveness in Section \ref{sec:comparativestudy}, looking at computational costs in Supplementary (Suppl.) Material, Part F.

\subsection{Notion of Data Depth}
\label{depthfunctions}
A data depth is a concept for measuring the centrality of a multivariate observation $\bm{m}_i$ (cf. \citealp{zuo2000general, liu2006data, mosler2022choosing}) with respect to a given reference sample $R_c  = \{ m_t :  = 1, \dots, |R|, y_t = c\}$, where $|R|$ defines its size which is the same for all considered classes. In other words, it creates a center-outward ordering of points in the Euclidean space of any dimension. There are various notions of data depth, each of them providing a distinctive center-outward ordering of sample points in a multidimensional space. In this work, we consider four data depth notions: Halfspace, Mahalanobis, Projection, and Simplicial depths.

First, the Halfspace depth (originally known as ``Tukey depth'') introduced by \cite{tukey1975mathematics} and further developed by \cite{donoho1992breakdown} is defined as the smallest number of data points in any closed halfspace with boundary hyperplane through $\bm{m}_i$ \citep{struyf1999halfspace}. That is,
\begin{equation*}
D_{H}^c(\bm{m}_i,\ R_c) = \frac{1}{|R|}\min_{\lVert \bm{p} \rVert = 1}|\{b:\ \langle\ \bm{p}, \bm{m}_b\rangle \geq \langle\ \bm{p},\bm{m}_i\rangle\}|,
\end{equation*}
where $|\cdot|$ denotes the cardinality of the set $B$ with $\bm{m}_b \in R_c$, $\bm{p}$ are all possible directions with $\lVert \bm{p} \rVert = \sqrt{\langle p,\ p \rangle}$ being the Euclidean norm and $\langle \cdot,\cdot \rangle$ the inner product. In our work, we consider its robust version \textbf{HD}$_r$ that is proposed by \cite{ivanovs2021distributionally} and calculated approximately, offering some advantages in being strictly positive and continuous beyond the convex hull of the observed samples.

Second, we consider the Mahalanobis depth which is based on the Mahalanobis distance (cf. \citealp{mahalanobis1936generalised}). It is derived as
\begin{equation*}
D_{M}^c(\bm{m}_i,\ R_c) = \dfrac{1}{1 + (\bm{m}_i - \bm{\mu}_{\bm{m}})' \bm{\Sigma}^{-1} (\bm{m}_i - \bm{\mu}_{\bm{m}})},
\end{equation*}
where $\bm{\mu}_{\bm{m}}$ is the mean vector of the embeddings in the reference sample and $\bm{\Sigma}^{-1}$ is the covariance matrix, estimated by the sample mean and the sample covariance matrix, respectively. 

Third, the Projection depth proposed by \cite{zuo2000general} is specified as
\begin{equation*}
D^c_P(\bm{m}_i,\ R_c) = \bigg(1 + \sup_{\lVert \bm{p} \rVert = 1}\dfrac{|\langle\bm{p},\bm{m}_i\rangle - \text{med}(\langle\bm{p}, R_c\rangle)|}{\text{MAD}(\langle\bm{p}, R_c\rangle)}\bigg)^{-1}
\end{equation*}
with $\langle\bm{p},\bm{m}_i\rangle$ denoting the inner product and the projection of $\bm{m}_i$ to $\bm{p}$ if $\lVert \bm{p} \rVert = 1$. The notation med($E$) defines the median of a univariate random variable $E$ and MAD($E$) $= \text{med}(|E - \text{med}(E)|)$ is the median absolute deviation from the median. As the exact computation of the Projection depth is possible only at very high computational costs (cf. \citealp{mosler2022choosing}), we use the algorithms that enable its calculation approximately. \cite{dyckerhoff2021approximate} provide the implementation and comparison of various algorithms. In our work, we study the performance of control charts based on three different algorithms to compute the Projection depth of both symmetric and asymmetric types. In particular, we consider coordinate descent (\textbf{PD}$_1$), Nelder-Mead (\textbf{PD}$_2$), and refined random search (\textbf{PD}$_3$) for the symmetric type, and \textbf{PD}$^a_1$,  \textbf{PD}$^a_2$, and \textbf{PD}$^a_3$ for the asymmetric type.

Fourth, we calculate the Simplicial depth \citep{liu1990notion} as
\begin{equation*}
D_{S}^c(\bm{m}_i, R_c) =  {\binom{|R|}{k + 1}}^{-1}\sum_{\diamond}I_{S(\bm{m}_t \, | \, t \in R_c, \ y_t = c)}(\bm{m}_i),
\end{equation*}
where $S(\bm{m}_t | t \in R_c, y_t = c)$ defines the open simplex consisting of vertices $\{\bm{m}_{t,1},$ $ \ldots, \bm{m}_{t,k+1}\}$ from all observations $t$ in the reference sample $R_c$. The $\diamond$ notation means that we validate all possible combinations to construct an open simplex with $(k + 1)$ vertices. We specify $I_A(x)$ as the indicator function on a set $A$ returning $1$ if $x \in A$ and $0$ otherwise. Both Simplicial (\textbf{SD}) and Mahalanobis (\textbf{MD}) depths are computed with algorithms provided by \cite{pokotylo2019depth}.  

Related to the classification problem of multivariate data, there exist depth-based classifiers (cf. \citealp{vencalek2017depth}) such as depth-vs-depth plot (DD-plot) designed by \cite{li2012dd} or DD-alpha procedure proposed by \cite{lange2014fast}. Also, the field of outlier or anomaly detection is a widespread area for data depth usage (cf. \citealp{dang2010nonparametric, baranowski2021transient}). Combining these two perspectives, we apply a data depth-based Shewhart-type $r$ control chart for single observations (see Sections \ref{rAndQ} and \ref{sec:comparativestudy}) and a batch-wise $Q$ control chart (see Suppl. Material, Part D) developed by \cite{liu1995control} for detecting nonstationarity in a data stream.

\subsection{The $r$ Control Chart}
\label{rAndQ}
A control chart is a graphical tool for monitoring processes by recording the performance of quality characteristics over time or sample number \citep{kan2002metrics}. The process is considered in control when the test statistic falls within the Upper and Lower Control Limits ($UCL$ and $LCL$). Points outside this range indicate unusual variability, triggering a signal to investigate the out-of-control state of the process.

\begin{figure}
	\begin{center}
		\includegraphics[width=1.00\textwidth]{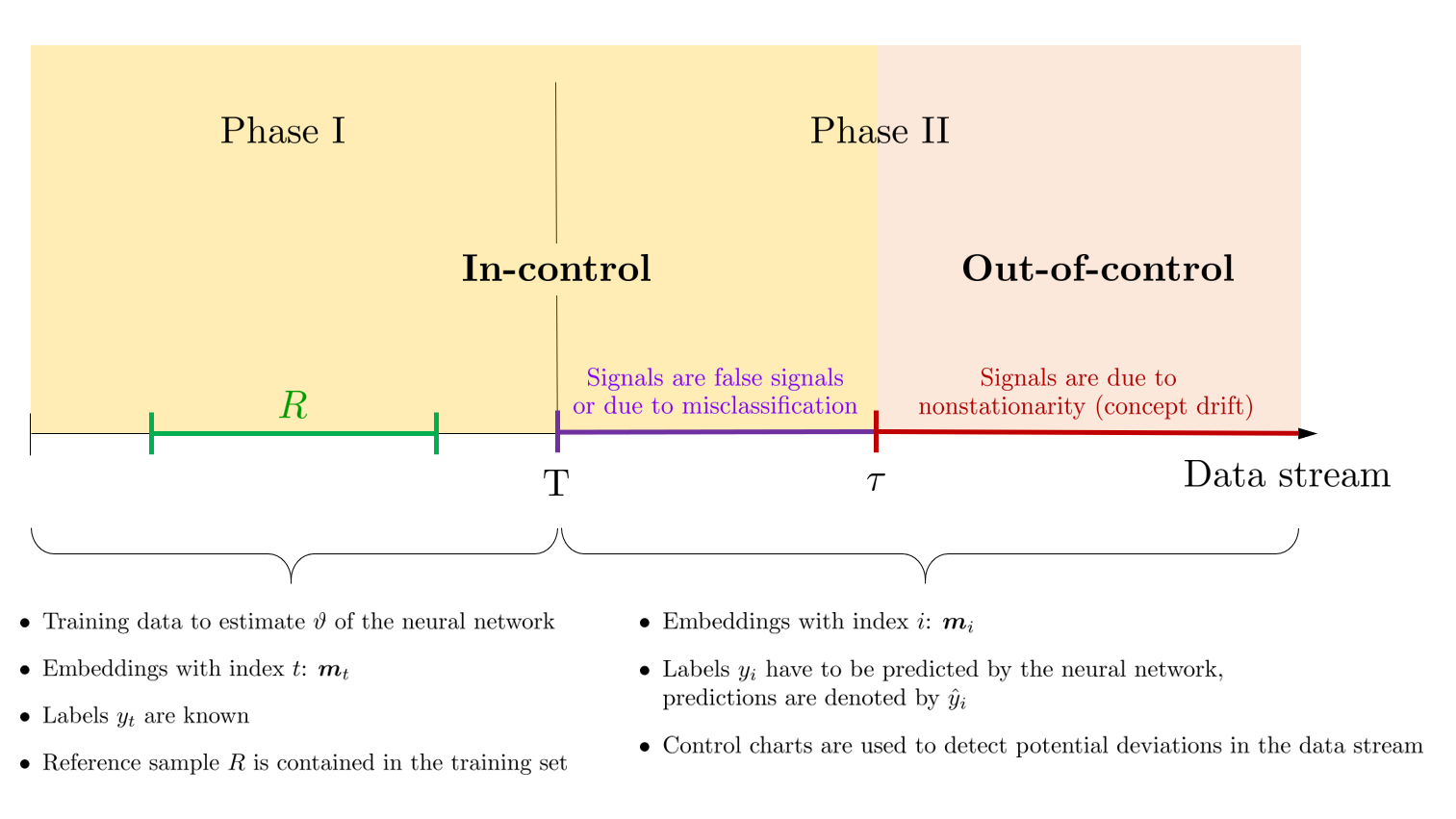} 
		\caption{Summary of the introduced notation and data subdivision.}
		\label{datasets}
	\end{center}
\end{figure}

Usually, the application of control charts is divided into Phase I and Phase II. In Phase I, we collect the reference data, examine its quality and verify the process stability, then estimate model parameters if applicable and derive the values for the control limits \citep{jones2014overview}. The data in Phase I does not have to coincide with the full training data of the ANN but could rather be its subset. That is, the sets $R_c \subseteq \{\bm{m}_t : t = 1, \ldots, T,\ y_t = c\}$ of size $|R|$ create the Phase I data where it is essential to consider only correctly classified data samples. Successively, in Phase II, the control chart statistic is plotted for each embedding $\bm{m}_i$ with $i > T$. Figure \ref{datasets} displays the introduced periods and sets. 

Considering the $r$ control chart proposed by \cite{liu1995control}, the scheme is based on ranks of multivariate observations, which are obtained by computing data depth. To determine whether $\bm{m}_i$ belongs to $\Xi_{c}$, we use the following control chart statistic 
\begin{equation*}
r_{\cdot}^c(\bm{m}_i) = \dfrac{|\{D_{\cdot}^c(\bm{m}_t) \leq D_{\cdot}^c(\bm{m}_i):  t \in R_c, \ y_t = c\}|}{|\{t\in R_c:\  y_t = c\}|}
\end{equation*}
that defines the rank of the observed depth related to the observations in the reference sample with a class $c$. Thus, the $r$ control chart monitors the values of $r_{\cdot}^c$ over time.
Considering the interpretation of ranks, we can state that $r_{\cdot}^c(\bm{m}_i)$ reflects how outlying $\bm{m}_i$ is with respect to the reference sample. If $r_{\cdot}^c(\bm{m}_i)$ is high, then there is a considerable proportion of data in the reference sample that is more outlying compared to $\bm{m}_i$ \citep{liu1995control}.

Regarding the control limits, there is no need to introduce the $UCL$ as $r_{\cdot}^c$ belongs to the continuous interval $[0, 1]$. Considering the $LCL$, it coincides with the significance level of the hypothesis test, here defined as $\alpha$. Thus, the process is considered to be out of control if $r_{\cdot}^c(\bm{m}_i) \leq \alpha$. The choice of $\alpha$ depends on the specification of Average Run Length ($ARL$) -- the expected number of monitored data points required for the control chart to produce a signal \citep{stoumbos2001nonparametric}. In the case of the Shewhart-type control charts, the reciprocal of $ARL$ corresponds to the false alarm rate ($FAR$) in the in-control state of a process. Technically speaking, since the $r$ control chart is a Shewhart control chart, $FAR = \alpha$, where $\alpha$ is interpreted as the probability of a false alarm in Phase I \citep{stoumbos2001nonparametric}. 

According to \cite{liu1995control}, the $r$ control chart can be applied with affine-invariant notions of data depth, explicitly mentioning Simplicial depth, Mahalanobis depth, and Halfspace depth. Since Projection depth is also affine-invariant (cf. \citealp{mosler2013depth}), both $r$ and $Q$ control charts can be combined with each of the data depth functions introduced in Section \ref{depthfunctions}.

\section{Comparative Study} 
\label{sec:comparativestudy}

In the following section, we analyze the effectiveness of the proposed monitoring framework by designing a toy example and using real data. The discussion begins with a description of the considered benchmark methods, followed by the introduction of the selected performance measures in Section \ref{sec:performance_measures}. Further, we compare the performances for toy example in Section \ref{ToyExample} and for real data in Section \ref{realdata}. In addition, the construction of reference samples and the misclassification effect are examined in Section \ref{subsec:misclassif}.

\subsection{Benchmark Methods}
\label{Benchmark}

Due to the independent development of comparable approaches (cf. \citealp{mozharovskyi2022anomaly, yang2021generalized}) and their focus on different scenarios/perspectives, there is no unified benchmark. Hence, to select a benchmark for ANN monitoring, we need to consider the available options. Three possibilities arise: inspecting the initial input, the model's embeddings, or the final output (softmax score in our case) represented by layers A, B, and C in Figure \ref{a:FNN}. Monitoring the initial data quickly becomes limited due to the complexity of typical datasets analyzed by ANNs. On the contrary, using intermediate layers reduces data dimensionality and storage requirements since only the embeddings from the training phase need to be saved. Therefore, we only consider options B and C as monitoring options. Furthermore, as a benchmark, we focus on methods that can operate within the introduced framework. Specifically, we seek methods capable of detecting nonstationarity in (1) individual samples (batch size of one), (2) without the need for labels in Phase II, and (3) working with or without available time stamps. To ensure comparability, we utilize the same SPM framework of $r$ control charts.

Based on the recent reviews (cf. \citealp{goldstein2016comparative, villa2021semi, yang2021generalized}), we choose a Kernel Density Estimation Outlier Score (\textbf{KDEOS}) defined by \cite{schubert2014generalized}, a distance-based Local Outlier Factor (\textbf{LOF}) developed by \cite{breunig2000lof} and an ensemble-based outlier detection method such as isolation Forest (\textbf{iForest}) proposed by \cite{liu2008isolation} as a common benchmark coming from distribution- and distance-based methods. Both \textbf{LOF} and \textbf{KDEOS} compare the densities within local neighborhoods. However, while \textbf{LOF} is based on the reachability distance of the point to its neighborhood for density estimation, \textbf{KDEOS} uses classic kernel density estimates, e.g., based on Gaussian or Epanechnikov kernels. The \textbf{iForest} represents an ensemble of binary decision trees, where the points placed deeper in the trees are less likely to be outliers as they require more splits of space to isolate them. On the contrary, the samples which are allocated in shorter branches would rather be anomalous. 

When considering option C, we compare our approach with monitoring the softmax scores. They are normalized between 0 and 1, and their length is equal to the number of neurons in the final layer. The neuron that has the maximum score corresponds to the predicted class. It is important to note that the softmax output is widely considered as a measure of the model's confidence (cf. \citealp{gawlikowski2021survey, moon2020confidence}); however, there is substantial research into the area of alleviating overconfident prediction issue \citep{gawlikowski2021survey}, {e.g.}, by redesigning a loss function that leads to more trustful confidence estimates \citep{moon2020confidence}. Nevertheless, directly using the softmax output for nonstationarity detection is considered to perform reasonably well \citep{pearce2021understanding}. Thus, as benchmark techniques, we select Mahalanobis distance (\textbf{MDis}) which is well-known for detecting concept drift in similar settings (cf. \citealp{lee2018simple, yang2021generalized}), and a Natural Outlier Factor (\textbf{NOF}) based on Natural Neighbour principle, where calculation of the factor is parameterless \citep{huang2016non}.

\subsection{Performance Measures}\label{sec:performance_measures}

A well-operating control chart has a low false alarm rate (when it signals a change incorrectly) and a high rate of correctly detected out-of-control points. The performance of control charts is typically assessed by $ARL$. It measures the time until a false alarm when the process is in control and the speed at which the chart detects an actual change when the process is out of control. Alternatively, the False Alarm Rate ($FAR$) evaluates performance in Phase I, while the Signal Rate ($SR$) and Correct Detection Rate ($CDR$) assess performance in Phase II. All three metrics range between 0 and 1, providing the relative number of false or correct signals.

Regarding the $SR$ value, we calculate a proportion of false alarms given the total length of the considered in-control part. In the case of the $CDR$ value, it is computed as a proportion of correctly detected out-of-control data points given the total length of the designed out-of-control part. To account for a possible discrepancy between the class proportions of the predicted data in Phase II, we calculate the weighted mean of the occurred signals, accounting for the number of data points in each predicted class within the observed period. In the case of $FAR$, we simply use the sample mean because the class or reference sample sizes are identical. 

If a tested control chart operates as desired, then $FAR$ of Phase I equals the chosen probability of a false alarm $\alpha$. In Phase II, ideally, we would expect $SR$ to be similar to $FAR$ for the in-control samples (neglecting the potential misclassification effect), while the $CDR$ should be as large as possible for the out-of-control samples. If the $CDR$ is low, i.e., close to 0, we conclude that the control chart does not accomplish its primary purpose  -- to detect nonstationarity in a data stream.

\subsection{Toy Example}
\label{ToyExample}

\begin{table}[t]
	\centering
	\scriptsize
		\renewcommand{\arraystretch}{0.8}
	\begin{tabular}{|c|c|c|c|c|cc:ccccccc|}
	\hline
	\multicolumn{1}{|c}{\textbf{Evaluation}} &
		\multicolumn{1}{c}{Size $|R|$} &
		\multicolumn{1}{c}{Phase} &
		\multicolumn{1}{c}{Observed process} &
		\multicolumn{1}{c}{Metric} &
		\multicolumn{1}{|c}{\textbf{MD}}  &
		\multicolumn{1}{c:}{\textbf{SD}} &
		\multicolumn{1}{c}{\textbf{HD$_r$}}  &
		\multicolumn{1}{c}{\textbf{PD$^a_1$}} &
		\multicolumn{1}{c}{\textbf{PD$^a_2$}} &
		\multicolumn{1}{c}{\textbf{PD$^a_3$}} &
		\multicolumn{1}{c}{\textbf{PD$_1$}} &
		\multicolumn{1}{c}{\textbf{PD$_2$}} &
		\multicolumn{1}{c|}{\textbf{PD$_3$}}\\
		\hline
		
\rowcolor{green!20}
 \cellcolor{white}& 100 & I & In-control & \textit{FAR} & 0.05 & 0.03 & 0.05 & 0.05 & \underline{0.05} & 0.05 & 0.05 & 0.05 & 0.05 \\ 
 \rowcolor{blue!15}
 \cellcolor{white} Toy example & 100 & II & In-control & \textit{SR} & 0.08 & 0.00 & 0.10 & 0.08 & \underline{0.04} & 0.06 & 0.06 & 0.08 & 0.10 \\ 
 \rowcolor{blue!15}
 \cellcolor{white} $\bm{m}_i \in \mathbb{R}^{3}$& 100 & II & Out-of-control & \textit{CDR} & 1.00 & 0.00 & 1.00 & 1.00 & \underline{1.00} & 1.00 & 0.10 & 0.06& 0.14 \\ 
\hline

		&\cellcolor{green!20}2000 &\cellcolor{green!20}I& \cellcolor{green!20} In-control&\cellcolor{green!20}$FAR$ &\cellcolor{green!20}0.05
		&\cellcolor{green!20}/
		&\cellcolor{green!20}0.05
		& \cellcolor{green!20}0.05
		&\cellcolor{green!20}0.05
		&\cellcolor{green!20}0.05
		& \cellcolor{green!20}0.05
		&\cellcolor{green!20}\underline{0.05}
		&\cellcolor{green!20}0.05 \\ 
		& \cellcolor{blue!15}2000 &\cellcolor{blue!15}II& \cellcolor{blue!15} In-control& \cellcolor{blue!15}$SR$ & \cellcolor{blue!15}0.51 & \cellcolor{blue!15}/& 
		\cellcolor{blue!15}0.54& 
		\cellcolor{blue!15}0.49& \cellcolor{blue!15}0.49&\cellcolor{blue!15}0.50&
		\cellcolor{blue!15}0.48& \cellcolor{blue!15}\underline{0.47}&
		\cellcolor{blue!15}0.48\\
	 \rowcolor{blue!05}
	\tiny{	\cellcolor{white}}&\tiny{}&\tiny{}&\tiny{}&\tiny{$SR|M$} & \tiny{0.97}&\tiny{/}& \tiny{0.98}& \tiny{0.96} & \tiny{0.97}& \tiny{0.97}& \tiny{0.96} & \tiny{0.96} & \tiny{0.96}\\
		 \rowcolor{blue!05}
	\tiny{	\cellcolor{white}}&\tiny{}&\tiny{}&\tiny{}&\tiny{$SR|C$} & \tiny{0.46}&\tiny{/}& \tiny{0.49}& \tiny{0.44} & \tiny{0.44}& \tiny{0.45}& \tiny{0.43} & \tiny{0.42} & \tiny{0.42}\\
		& \cellcolor{blue!15}2000	&\cellcolor{blue!15}II& \cellcolor{blue!15} Out-of-control
		&\cellcolor{blue!15} $CDR$
		& \cellcolor{blue!15}0.92 &\cellcolor{blue!15}/
		&\cellcolor{blue!15}0.93
		& \cellcolor{blue!15}0.92
		& \cellcolor{blue!15}0.92
		&\cellcolor{blue!15}0.92
		& \cellcolor{blue!15}0.91
		& \cellcolor{blue!15}\underline{0.91}
		&\cellcolor{blue!15}0.89 \\
		\hhline{~-------------}
 	Experiment 1 &\cellcolor{green!20}3000 	&\cellcolor{green!20}I& \cellcolor{green!20} In-control&\cellcolor{green!20}$FAR$ &\cellcolor{green!20}0.05
		& \cellcolor{green!20}/
		&\cellcolor{green!20}0.05
		& \cellcolor{green!20}0.05
		&\cellcolor{green!20}0.05
		&\cellcolor{green!20}0.05
		& \cellcolor{green!20}0.05
		&\cellcolor{green!20}\underline{0.05}
		&\cellcolor{green!20}0.05 \\ 
		$\bm{m}_i \in \mathbb{R}^{16}$& \cellcolor{blue!15}3000	&\cellcolor{blue!15}II& \cellcolor{blue!15} In-control& \cellcolor{blue!15}$SR$  & \cellcolor{blue!15}0.43 & \cellcolor{blue!15}/&
		\cellcolor{blue!15}0.46&
		\cellcolor{blue!15}0.40& \cellcolor{blue!15}0.41&
		\cellcolor{blue!15}0.41&\cellcolor{blue!15}0.41& \cellcolor{blue!15}\underline{0.39}&
		\cellcolor{blue!15}0.39\\
		\rowcolor{blue!05}
	\tiny{	\cellcolor{white}}&\tiny{}&\tiny{}&\tiny{}&\tiny{$SR|M$} & \tiny{0.95}&\tiny{/}& \tiny{0.95}& \tiny{0.92} & \tiny{0.92}& \tiny{0.93}& \tiny{0.93} & \tiny{0.92} & \tiny{0.92}\\
		 \rowcolor{blue!05}
	\tiny{	\cellcolor{white}}&\tiny{}&\tiny{}&\tiny{}&\tiny{$SR|C$} & \tiny{0.38}&\tiny{/}& \tiny{0.40}& \tiny{0.35} & \tiny{0.35}& \tiny{0.35}& \tiny{0.35} & \tiny{0.33} & \tiny{0.33}\\
		& \cellcolor{blue!15}3000			&\cellcolor{blue!15}II& \cellcolor{blue!15} Out-of-control
		&\cellcolor{blue!15} $CDR$
		& \cellcolor{blue!15}0.88 &\cellcolor{blue!15}/
		&\cellcolor{blue!15}0.90
		& \cellcolor{blue!15}0.86
		& \cellcolor{blue!15}0.87
		&\cellcolor{blue!15}0.86 
		& \cellcolor{blue!15}0.87
		& \cellcolor{blue!15}\underline{0.86}
		&\cellcolor{blue!15}0.84 \\
		
		\hhline{~-------------}
		&\cellcolor{green!20}4000	&\cellcolor{green!20}I& \cellcolor{green!20} In-control&\cellcolor{green!20}$FAR$  &\cellcolor{green!20}0.05
		& \cellcolor{green!20}/
		&\cellcolor{green!20}0.05
		& \cellcolor{green!20}0.05
		&\cellcolor{green!20}\underline{0.05}
		&\cellcolor{green!20}0.05
		& \cellcolor{green!20}0.05
		&\cellcolor{green!20}0.05
		&\cellcolor{green!20}0.05 \\ 
		& \cellcolor{blue!15}4000 &\cellcolor{blue!15}II& \cellcolor{blue!15} In-control& \cellcolor{blue!15}$SR$ & \cellcolor{blue!15}0.34 & \cellcolor{blue!15}/& 
		\cellcolor{blue!15}0.37 & \cellcolor{blue!15}0.32& \cellcolor{blue!15}\underline{0.31}&
		\cellcolor{blue!15}0.31&\cellcolor{blue!15}0.33& \cellcolor{blue!15}0.31&
		\cellcolor{blue!15}0.30\\
		\rowcolor{blue!05}
	\tiny{	\cellcolor{white}}&\tiny{}&\tiny{}&\tiny{}&\tiny{$SR|M$} & \tiny{0.88}&\tiny{/}& \tiny{0.91}& \tiny{0.83} & \tiny{0.84}& \tiny{0.83}& \tiny{0.86} & \tiny{0.83} & \tiny{0.83}\\
		 \rowcolor{blue!05}
	\tiny{	\cellcolor{white}}&\tiny{}&\tiny{}&\tiny{}&\tiny{$SR|C$} & \tiny{0.29}&\tiny{/}& \tiny{0.32}& \tiny{0.26} & \tiny{0.26}& \tiny{0.26}& \tiny{0.27} & \tiny{0.25} & \tiny{0.25}\\
		& \cellcolor{blue!15}4000 &\cellcolor{blue!15}II& \cellcolor{blue!15} Out-of-control
		&\cellcolor{blue!15} $CDR$
		& \cellcolor{blue!15}0.79 &\cellcolor{blue!15}/
		&\cellcolor{blue!15}0.83
		& \cellcolor{blue!15}0.78
		& \cellcolor{blue!15}\underline{0.79}
		&\cellcolor{blue!15}0.78
		& \cellcolor{blue!15}0.78
		& \cellcolor{blue!15}0.78
		&\cellcolor{blue!15}0.73 \\
		\hline	
		
	\end{tabular}
\caption[Performance of $r$ control charts ($\alpha = 0.05$) in toy example and in Experiment 1.]{Performance of $r$ control charts ($\alpha = 0.05$) in toy example and in Experiment 1 with reference samples $R$ being predicted classes. The underlined numbers indicate the suggested method for an entire monitoring period, based on the trade-off between $SR$ and $CDR$. We compute \textbf{SD} for $\mathbb{R}^{3}$ only, as computational complexity is $O(n^{k+1})$.}
	\label{ToyExample:Depths}
\end{table}

To present our idea in a controllable environment, we create a toy example using the ANN architecture presented in Figure \ref{a:FNN}. We simulate two 7-dimensional Gaussian random variables $\bm{x}_t^{(1)}$ and $\bm{x}_t^{(2)}$ with $\bm{\mu}_1 = \bm{0}$ and $\bm{\mu}_2 = 10\cdot \bm{1}_7$, where $t = 1, \dots, 100$ and $\bm{1}_n$ is the $n$-dimensional vector of ones. Both variance-covariance matrices $\bm{\Sigma}_1$ and $\bm{\Sigma}_2$ have $\sigma_{ii} = 1$ but with $\sigma_{i-1, j} = \sigma_{i, j-1} = 0.3$ in the case of $\bm{\Sigma}_1$, and $\sigma_{i-1, j} = \sigma_{i, j-1} = -0.3$ in case of $\bm{\Sigma}_2$ for all $i,j = 1,\dots,7$ (in respective cases $i, j > 1$), where the remaining entries are zero. Considering the out-of-control data, we sample from a new multivariate Gaussian distribution with $\bm{\mu}_\tau = 5\cdot\bm{1}_7$ and $\bm{\Sigma}_1$.

We use a reference sample of size $|R| = 100$ for each of the classes, i.e., the entire training data because there were no misclassified data points. In Phase II, the in-control data corresponds to 50 new observations with the same distribution as used for training, while the out-of-control 50 data points correspond to the out-of-control distribution. The embedding layer consists of three neurons, i.e. $\bm{m}_i \in \mathbb{R}^{3}$. The visualization of the embeddings that correspond to both reference samples and out-of-control data is displayed in Figure \ref{b:vis}.

\begin{table}[t]
	\centering
	\scriptsize
	\renewcommand{\arraystretch}{0.80}
	\begin{tabular}{|c|c|c|c|c|cc|ccc||c|}
	\hline
	 \multicolumn{5}{|c|}{} & 
	 \multicolumn{2}{c|}{$\bm{m}_i \in \mathbb{R}$} & 
	  \multicolumn{3}{c||}{$\bm{m}_i \in \mathbb{R}^{16}$}&
	  	  \multicolumn{1}{c|}{}\\
	  \multicolumn{1}{|c}{\textbf{Evaluation}} &
	  \multicolumn{1}{c}{Size $|R|$} &
		\multicolumn{1}{c}{Phase} &
		\multicolumn{1}{c}{Observed process} &
		\multicolumn{1}{c|}{Metric} &
		\multicolumn{2}{c|}{}&
		\multicolumn{3}{c||}{}&
		\multicolumn{1}{c|}{}\\
		\multicolumn{5}{|c|}{}  &
		\multicolumn{1}{c}{\textbf{MDis}}  &
		\multicolumn{1}{c|}{\textbf{NOF}} &
		\multicolumn{1}{c}{\textbf{KDEOS}} &
		\multicolumn{1}{c}{\textbf{LOF}} &
		\multicolumn{1}{c||}{\textbf{iForest}}&
		\multicolumn{1}{c|}{\textbf{PD$_2^{(a)}$}}\\
		\hline
		\rowcolor{green!20}

\cellcolor{white} & 100 & I & In-control & \textit{FAR} & 0.05 & \underline{0.05} & 0.05 & \underline{0.05} & 0.05 & \underline{0.05}\\ 
\rowcolor{blue!15}
\cellcolor{white}Toy example & 100 & II & In-control & \textit{SR} & 0.08 & \underline{0.04} & 0.00 & \underline{0.04}  & 0.12 & \underline{0.04}\\ 
\rowcolor{blue!15}
\cellcolor{white} & 100 & II & Out-of-control & \textit{CDR} & 1.00 & \underline{1.00}  & 1.00  & \underline{1.00} & 1.00 & \underline{1.00}\\

	\hline
	\rowcolor{green!20}
\cellcolor{white}  & 2000 & I & In-control & \textit{FAR} & 0.05 &  0.05& 0.05 & 0.05 &  0.05 & \underline{0.05}\\ 
\rowcolor{blue!15}
\cellcolor{white}& 2000 & II & In-control & \textit{SR} & 0.57 & 0.60& 0.07  & 0.56 & 0.53 & \underline{0.47}\\ 
\rowcolor{blue!15}
\cellcolor{white} &2000 & II & Out-of-control & \textit{CDR} & 0.92 & 0.92& 0.05 & 0.93 & 0.93 & \underline{0.91}\\ 
	\hhline{~----------}
		\rowcolor{green!20}
	\cellcolor{white}  & 3000 & I & In-control & \textit{FAR} & 0.05& 0.05& 0.05 & 0.05 & 0.05  & \underline{0.05}\\ 
\rowcolor{blue!15}
\cellcolor{white}Experiment 1& 3000 & II & In-control & \textit{SR} & 0.44 & 0.47& 0.07 & 0.48 & 0.46 & \underline{0.39}\\ 
\rowcolor{blue!15}
\cellcolor{white} &3000 & II & Out-of-control & \textit{CDR} & 0.87 &0.86& 0.07 & 0.87 & 0.89 & \underline{0.86}\\ 	
\hhline{~----------}		
	\rowcolor{green!20}
	\cellcolor{white}  & 4000 & I & In-control & \textit{FAR} & 0.05 &0.05& 0.05 & 0.05 & 0.05 & \underline{0.05}\\ 
\rowcolor{blue!15}
\cellcolor{white}& 4000 & II & In-control & \textit{SR} & 0.33 &0.36& 0.07 & 0.39 & 0.38 & \underline{0.31}\\ 
\rowcolor{blue!15}
\cellcolor{white} &4000 & II & Out-of-control & \textit{CDR} & 0.79&0.78& 0.10 & 0.77 & 0.83 & \underline{0.79}\\ 	
\hline
	\end{tabular}
\caption[Comparison study: Performance of $r$ control charts ($\alpha = 0.05$) in toy example and in Experiment 1.]{Comparison study: Performance of $r$ control charts ($\alpha = 0.05$) in toy example and in Experiment 1. In the case of \textbf{MDis} and \textbf{NOF}, the data points represent the model's softmax output ($\bm{m}_i \in \mathbb{R}$). The underlined numbers indicate the suggested method based on the trade-off between $SR$ and $CDR$.}
\label{BenchmarkCIFAR}
\end{table}

Table \ref{ToyExample:Depths} summarizes the results from depth-based control charts. As we can see, all versions apart from symmetric Projection and Simplicial depths can be successfully applied in this setting. The reason why symmetric Projection depths fail is the asymmetric distribution of the processed data (see Figure \ref{b:vis}). Regarding the Simplicial depth, there are 24\% of data points in the reference sample of class 2 with $\textbf{SD}(\bm{m}_t)=0$. That can be explained by Simplicial depth assigning zero to every point in the space outside the sample's convex hull \citep{francisci2019generalization}. Moreover, because all out-of-control samples received the predictions of class 2, the \textbf{SD} could not detect the out-of-control samples. Comparing the best result from Table \ref{ToyExample:Depths} which is \textbf{PD$^a_2$} to the benchmark in Table \ref{BenchmarkCIFAR}, we notice that the \textbf{LOF} and \textbf{NOF} achieved similar results, while the \textbf{KDEOS} and \textbf{iForest} did not hold their size of $\alpha = 0.05$ in Phase II.

\subsection{Experiments with Real Data}
\label{realdata}
We conduct in total three experiments. In decreasing complexity, the first experiment is about a ten-class classification of images, applying Convolutional ANN (CNN), followed by a four-class classification of questions, using ANN with a Long Short-Term Memory layer (LSTM) and finished with a binary classification of sonar data performed with an FNN. Table \ref{Summary} provides a summary of the conducted experiments. The models' training aims to maximize overall classification accuracy, respectively tuning the hyperparameters and the ANN architectures. For the sake of brevity, we only report the results of Experiment 1 below and include the results of the other two experiments in Suppl. Material. 
\begin{table}
	\centering
	\small
	\begin{tabular}{p{7cm} ccc}
		
		\hline
	\mbox{}\\[-0.5cm]
		Experiment                                    &   1      &     2                &   3       \\
		\hline
	   Complexity                                     & High     &  Medium                 &    Low \\
       Data type                                      & Image    &   Sentence  & Signal in $[0,1]$    \\
       Type of NN                                     & CNN      &        LSTM           & FNN       \\
       Number of classes                              &   10     &      4            &       2       \\
       Phase I, in-control (Training data)            &   50000  &         2800       &        170   \\
       Phase II, in-control (Test data)               &   10000  &         600         &       38    \\
       Phase II, out-of-control (Nonstationary data)  &   400    &            60      &      30      \\
		\hline
	\mbox{}\\[-0.5cm]
       Results   &   Section \ref{sec:comparativestudy}  &      Suppl. Material, Part \ref{Experiment2}              &   Suppl. Material, Part \ref{Experiment3}    \\
		\hline
	\end{tabular}
	\caption[Summary of the experiments.]{Summary of experiments: Data size indicates the total number of samples across all classes. Balanced datasets are used for training the ANN, ensuring an equal representation of samples from each class.}
	\label{Summary}
\end{table}

\subsubsection{Multiclass Classification of Images}
\label{CIFAR}

\begin{figure}
	\begin{center}
		\includegraphics[width=0.9\textwidth, trim= 2.0cm 14.2cm 2cm 3cm,clip]{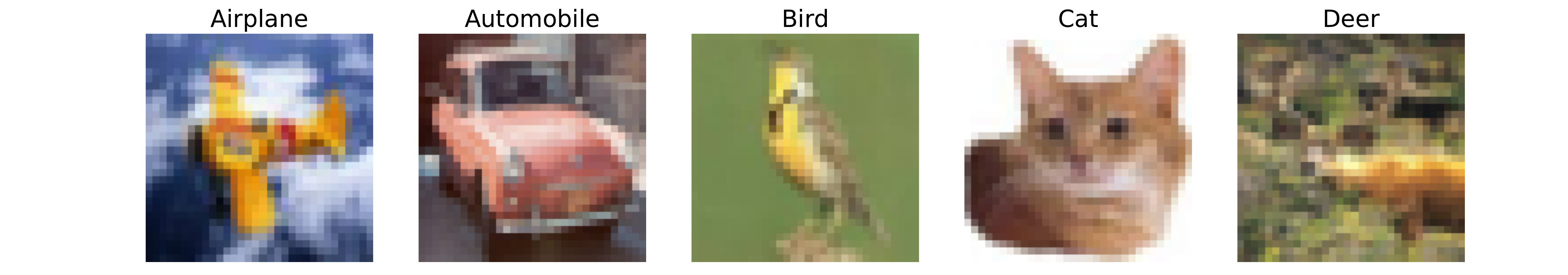} 
		\includegraphics[width=0.9\textwidth, trim= 2.0cm 3cm 2cm 13.6cm,clip]{CIFAR10.pdf} 
		\caption{Image examples and class labels of the CIFAR-10 dataset.}
		\label{CIFAR10}
	\end{center}
\end{figure}

In this experiment, we work with the CIFAR-10 dataset\footnote{\label{CIFARsource}\url{https://www.cs.toronto.edu/~kriz/cifar.html}} containing color images of the size  32$\times$32 pixels \citep{krizhevsky2009learning}, which is often applied for testing new out-of-distribution detection methods (cf. \citealp{yang2022openood}). In total, there are 60000 images which correspond to 6000 pictures per class. Figure \ref{CIFAR10} shows examples of each category. For the out-of-control samples, we consider the CIFAR-100 dataset\footnotemark[1] that has 100 image groups, selecting four distinctive classes, namely ``Kangaroo'', ``Butterfly'', ``Train'' and ``Rocket''. From each category we randomly chose 100 images, having in total 400 samples for the out-of-control part in Phase II. 

To construct a classifier for predicting to which of the ten groups an input image belongs, we train a CNN with a deep layer aggregation structure as proposed by \cite{yu2018deep}. A specification of such architecture is a tree-structured hierarchy of operations to aggregate the extracted features from different model stages. For a detailed introduction to CNNs, we direct to \cite{o2015introduction}. The embedding layer has 16 neurons, so we obtain a monitoring task of $ \bm{m}_i \in \mathbb{R}^{16}$. Regarding the training results after 88 epochs, the achieved accuracy on the test images was 90.43\%.

\subsubsection{Choice of Reference Samples and Monitoring Results}

\begin{figure}
	\begin{center}
		\includegraphics[width=0.65\textwidth, trim= 0.0cm 0.2cm 1cm 2cm,clip]{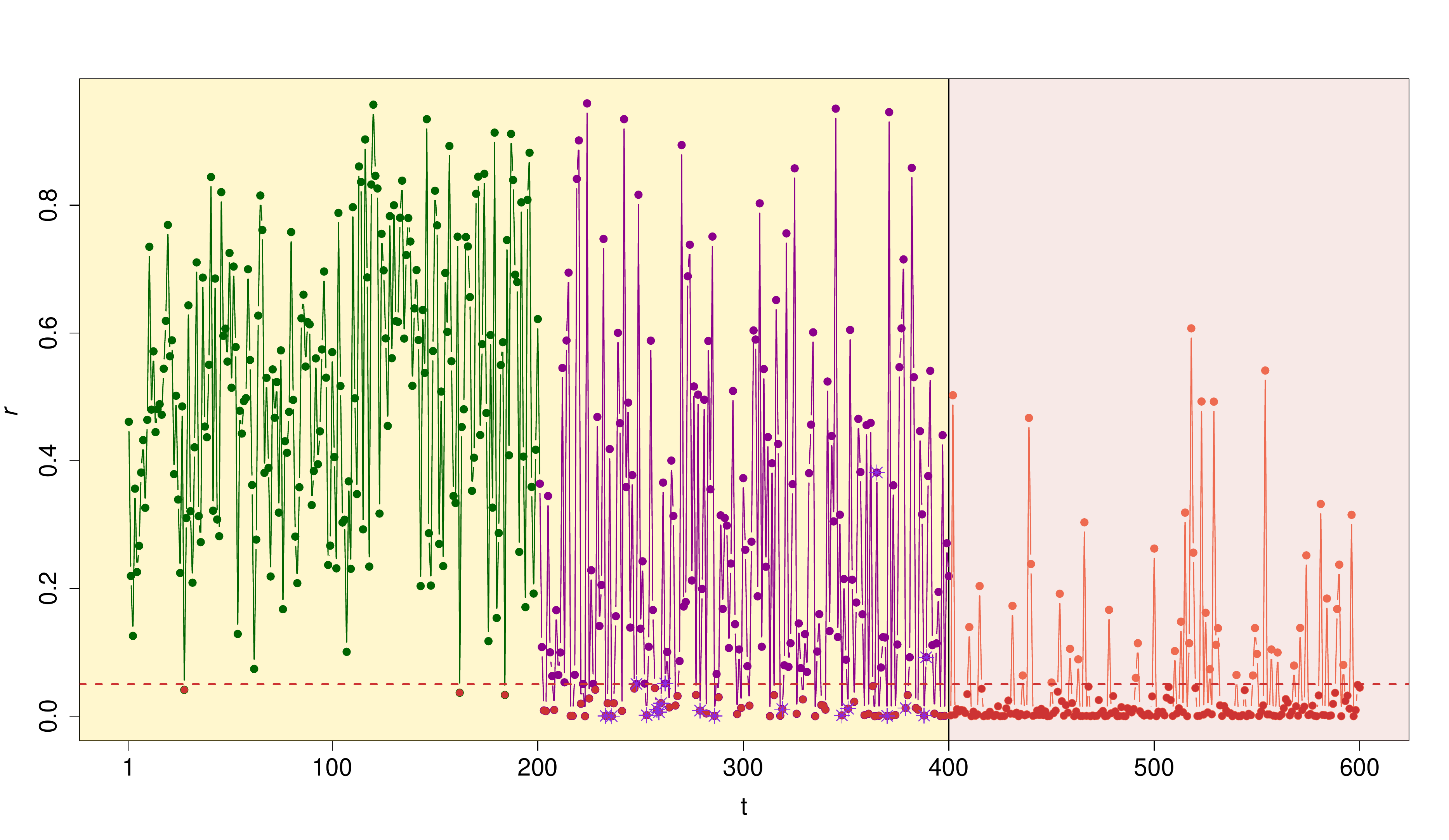} 
		\caption{An example of the $r$ control chart based on \textbf{PD}$_2$ using the data from Experiment 1. Colors correspond to the phases illustrated in Figure \ref{datasets}. The dashed line represents the control limit $\alpha = 0.05$, the signals are shown in dark red, and the misclassified samples in Phase II (in-control) are indicated with an asterisk.}
		\label{ControlChart}
	\end{center}
\end{figure}

Below, we investigate the effect of different reference samples, particularly concentrating on their size. To guarantee a well-chosen reference sample for each class, we construct it by choosing $|R|$ data points that obtained the highest softmax scores. The analysis of applying randomly created reference samples is provided in Suppl. Material, Part A. To illustrate the application of the \textit{r} control chart in Figure \ref{ControlChart}, we depict \textbf{PD}$_2$ with $|R| = 4000$. While we have 3 signals in Phase I (green points), where $FAR = 5\%$, a considerably larger number of signals is observed in Phase II without novelty (i.e., in-control, purple points). A substantial part of these signals occurred on the misclassified samples, marked by the asterisks. In the out-of-control part in Phase II (red points), we notice the highest number of signals, signifying correctly detected nonstationarity.

Looking at the results shown in Table \ref{BenchmarkCIFAR}, we can recognize the following patterns: First, with increasing reference sample size, the number of false alarms in Phase II decreases. For instance, \textbf{MD} and \textbf{HD}$_r$ lead to $SR$ being over 50\% when $|R| = 2000$, but we observe improvements by over 15\% when the size of the reference sample increases. Second, the larger the reference sample, the less precise becomes the out-of-control detection. Here, the $CDR$ values remain moderately high while doubling the size of the reference samples. Hence, it is beneficial to agree on such a reference sample size that slightly decreases $CDR$ and, at the same time, improves the performance of the control chart during the in-control state. With this strategy, both \textbf{PD$^a_2$} and \textbf{PD$_2$} suit the entire monitoring period well.

Similar behavior can be noticed for the benchmark methods presented in Table \ref{BenchmarkCIFAR}. Comparing the best depth-based monitoring result of \textbf{PD}$_2^{(a)}$ with the benchmark, we note that it outperforms all other algorithms. Nevertheless, the $SR$ values remain generally high, which could be due to misclassified observations being flagged as anomalous samples or because the dispersion of the test data is large compared to the reference samples. To better understand these issues, we analyze the data in Phase II more closely in the subsequent part.

\subsubsection{Misclassification and Data Diagnostic}
\label{subsec:misclassif}
To investigate the effect of misclassification, we refer to the wrongly classified images from the test data (Phase II, in-control), which constitute 958 data points. By calculating the signal rates conditional on misclassified $SR|M$ and correctly classified samples $SR|C$  for each depth notion and reference sample size in Table \ref{BenchmarkCIFAR}, we find out that the average $SR|M$ is 91.58\%. At the same time, $SR|C$ values are considerably lower and approach 25\% for bigger reference samples, compared to the original $SR$ results.

\begin{figure}
	\begin{center}
\includegraphics[width=0.70\textwidth, trim= 2.5cm 2cm 0.8cm 1.0cm,clip]{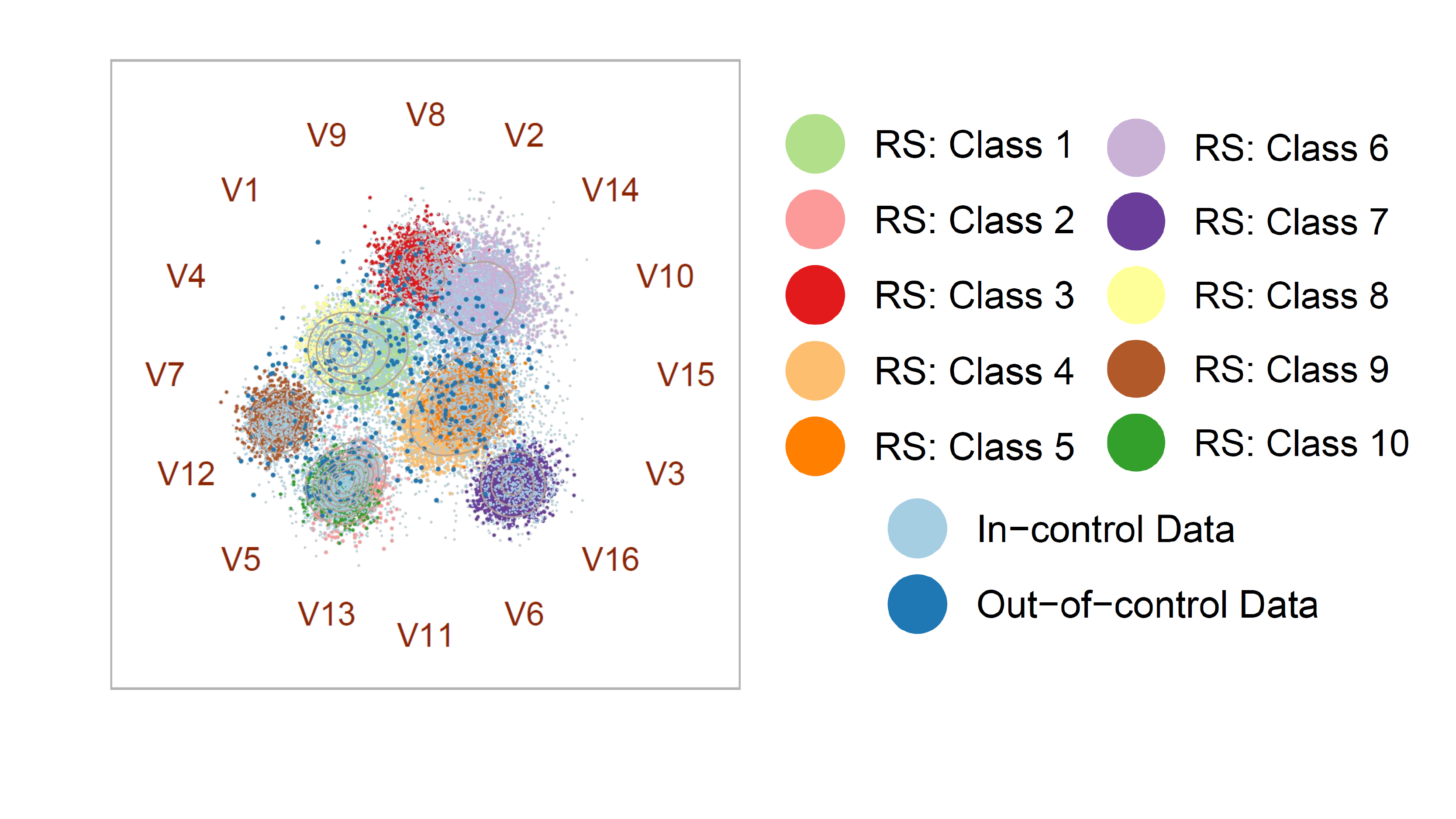} 
\caption{Visualization of the data from Experiment 1 with $|R| = 2000$. $V_1, \dots, V_{16}$ define anchors which correspond to neurons that produced embeddings $\bm{m}_i \in \mathbb{R}^{16}$.  Density contour plots outline respective classes.}
		\label{Plot16D}
	\end{center}
\end{figure}

To investigate another reason for high $SR$ values, we visualize the in-control data in Figure \ref{Plot16D}. We use Radial Coordinate visualization (Radviz), where the variables are referred to as anchors, being evenly distributed around a unit circle (cf. \citealp{hoffman1999dimensional, caro2010analyzing, abraham2017xploring}). Their order is optimized to place highly correlated variables next to each other. Correspondingly, data points are projected to positions close to the variables that have a higher influence on them. As we can see in Figure \ref{Plot16D}, there is a comparably large section opposite the anchor $V_{15}$ where the test data does not overlap with any of the reference samples. At the same time, a fraction of the out-of-control samples is located within reference samples, leading to more challenging detection. Hence, this analysis and the evaluation of the misclassification effect could facilitate the understanding of $SR > FAR$ and provide insight into $CDR$.

A possible solution for mitigating the misclassification effect and challenges in choosing a representative reference sample for each class could be a creation of a merged reference sample. In other words, we could neglect the condition of having class-specified reference samples and perform the computation of depths with respect to a grouped reference sample only. However, according to the results presented in Suppl. Material, Part E, such construction of reference samples could eliminate misclassification and sample selection issues but work only for less complex cases. Moreover, the computation time would increase rapidly for high-dimensional problems (see Suppl. Material, Part F), meaning that the proposed framework of using individual reference samples for each class would also be more suitable from this perspective.

\subsection{Practical Recommendations}

Overall, we notice that the asymmetric Projection depth works most reliably among the examined depths. The reason for that is twofold: First, it considers the geometry of the points, i.e., their asymmetric positioning. Second, as we usually aim to diagnose the outlyingness of the points that are outside of the convex hull, we need to be able to order them. By obtaining positive depth values outside the convex hull, we can better recognize which points are anomalous. On the contrary, the Simplicial or symmetric Projection depths underperform, if many points are placed outside the convex hull (an issue for \textbf{SD}) or the data is asymmetrically spread (an issue for \textbf{PD}).

As soon as a change has been detected, various actions could be implemented. \cite{lu2018learning} introduce the idea of ``Concept Drift Understanding'' before its adaptation. They stress that it is vital to answer how severe and in which data region the concept drift occurred before implementing further actions. Afterward, the model can be either adjusted or rebuilt, resulting in a new cycle of training and validation.

\section{Conclusion}\label{sec:conclusion}

The models based on ANNs have contributed to recent advances in various disciplines. However, the impressive results that were achieved with their deployment mask the necessity of the model's control and monitoring, meaning that critical flaws could occur without any notice by the expert.

This work proposes a monitoring procedure designed for ANN applications that applies a nonparametric multivariate control chart based on ranks and data depths. The core idea is to monitor the low-dimensional representation of input data called embeddings that are generated by ANNs. The proposed monitoring methodology has great potential and often outperforms benchmark methods in realistic experiments. Comparing different data depth notions under the trade-off between computation time and monitoring effectiveness, we recommend the asymmetric Projection depth using the Nelder-Mead algorithm. In case the embeddings are scattered symmetrically, symmetric Projection depth can be used instead.

Furthermore, we investigated the influence of the reference sample size and the method of how it was constructed. It could be shown that a bigger reference sample is not automatically related to a better detection capability. In particular, there are open questions about attaining a reliable reference sample. It is advisable to research in more detail how the SPM techniques, such as the multivariate mean-rank chart \citep{bell2014distribution}, could support the analysis of Phase I data. Also, it is subject to future research when the reference sample should be updated or continually augmented with recent observations while preventing contamination with the out-of-control data.

In our experiments, the data points of the training and the test datasets are chosen following a convention by randomly dividing the dataset. In the future, we recommend examining whether an optimal splitting of data into training and testing, which preserves distributional similarity, improves the performance of models based on AI as well as leads to more reliable monitoring (cf. \citealp{vakayil2022data}). Additionally, the field of data splitting and data compression, i.e., how to find a trade-off between reliable but fast training and a well-chosen training set that uses only a fraction of the initial dataset, is relevant for future research.

We have also shown that our approach would achieve a better performance if additional misclassification information were available. In general, understanding when a data point has obtained a wrong prediction is indispensable and requires an additional method to be developed that could be later combined with our monitoring procedure. Moreover, one could investigate whether subdividing the data stream in moving windows could enhance the monitoring performance.

In the empirical study, we consider well-balanced classification problems. However, class imbalance is a frequent challenge in training ANNs and needs to be considered in future research (cf. \citealp{ghazikhani2013ensemble}). Whether the proposed methodology could be applied to monitoring semi-supervised or unsupervised learning models remains also open. Despite of challenges in designing a universal monitoring scheme for AI-based approaches, extending the presented framework to other AI algorithms seems to be a promising field. Additionally, there are different types of concept drift or nonstationarity (cf. \citealp{hu2020no}), meaning that further development of methods and their comparison is of practical importance.

In summary, choosing a particular control chart depends on the specific problem, requiring a compromise between the number of signals in Phase II (In-control) and correctly detected out-of-control samples. As soon as one concludes what is important, namely computation time, robustness, or low variance, the proposed monitoring approach can be customized to satisfactorily support applications involving ANN.


\section*{Appendix}
\appendix

\section{Addition to Experiment 1: Monte Carlo Study (Choice of Reference Samples)}
To examine the influence of how the reference samples are formed, we conducted a Monte Carlo study with the data from Experiment 1. Here, we create reference samples for each class by randomly picking data points from correctly classified training data without considering confidence-related outcomes of the ANN. 

The obtained results based on 10 runs are summarized in Table \ref{MonteCarlo}. Due to the extensive computational resources involved, we conduct the study for one type of Projection depth, namely for the symmetric case.  For each choice of the data depth notion, the standard deviation does not exceed 0.01, meaning that the small number of iterations is sufficient for our investigation. The control charts based on \textbf{HD} achieve the highest $CDR$ among all proposed control charts. Furthermore, the control charts based on \textbf{PD}$_2$ and \textbf{PD}$_3$ are more reliable during Phase II (In-control), resulting in a $SR$ of 0.18. 

In contrast to the results where the reference samples are selected according to the softmax scores, we observe low fluctuation in performance with the changing size $|R|$ and lower $CDR$ values. Thus, we recommend choosing the reference sample based on the intended purpose of the monitoring. When accepting higher signal rates in the in-control phase (potentially due to misclassification), reference samples should be chosen based on the softmax scores. In turn, this leads to more sensitive detection of out-of-control samples. 

\begin{table}
	\centering
	\scriptsize
	\renewcommand{\arraystretch}{0.8}
	\begin{tabular}{|c|c|c|c|c|cc:cccc|}
		
		\hline
		\multicolumn{1}{|c}{\textbf{Evaluation}} &
		\multicolumn{1}{c}{Size $|R|$} &
		\multicolumn{1}{c}{Phase} &
		\multicolumn{1}{c}{Observed process} &
		\multicolumn{1}{c|}{Metric} &
		\multicolumn{1}{|c}{\textbf{MD}}  &
		\multicolumn{1}{c:}{\textbf{SD}} &
		\multicolumn{1}{c}{\textbf{HD$_r$}}  &
		\multicolumn{1}{c}{\textbf{PD$_1$}} &
		\multicolumn{1}{c}{\textbf{PD$_2$}} &
		\multicolumn{1}{c|}{\textbf{PD$_3$}}\\
		
		\hline
		
		&\cellcolor{green!20}2000& \cellcolor{green!20} I&\cellcolor{green!20}In-control &\cellcolor{green!20}$FAR$
		  &\cellcolor{green!20}0.05
		& \cellcolor{green!20}/
		&\cellcolor{green!20}\underline{0.05}
		& \cellcolor{green!20}0.05
		&\cellcolor{green!20}0.05
		&\cellcolor{green!20}0.05\\ 
		&\cellcolor{blue!15}2000& \cellcolor{blue!15}II & \cellcolor{blue!15}In-control & \cellcolor{blue!15}$SR$
		 &\cellcolor{blue!15}0.21
		& \cellcolor{blue!15}/
		&	\cellcolor{blue!15}\underline{0.23}& \cellcolor{blue!15}0.19
		&\cellcolor{blue!15}0.18&
		\cellcolor{blue!15}0.18\\
		&\cellcolor{blue!15}2000& \cellcolor{blue!15}II
		&\cellcolor{blue!15}Out-of-control 
		& \cellcolor{blue!15}$CDR$
		 &\cellcolor{blue!15}0.62
		& \cellcolor{blue!15}/
		&\cellcolor{blue!15}\underline{0.64}
		& \cellcolor{blue!15}0.58
		& \cellcolor{blue!15}0.59
		&\cellcolor{blue!15}0.59\\
		\hhline{~----------}
		&\cellcolor{green!20}3000& \cellcolor{green!20}I&\cellcolor{green!20}In-control&\cellcolor{green!20}$FAR$
		&\cellcolor{green!20}0.05
		& \cellcolor{green!20}/
		&\cellcolor{green!20}\underline{0.05}
		& \cellcolor{green!20}0.05
		&\cellcolor{green!20}0.05
		&\cellcolor{green!20}0.05 \\ 
		Experiment 1	&\cellcolor{blue!15}3000& \cellcolor{blue!15}II& \cellcolor{blue!15}In-control& \cellcolor{blue!15}$SR$ 
		&\cellcolor{blue!15}0.21
		& \cellcolor{blue!15}/
		&\cellcolor{blue!15}\underline{0.23}& \cellcolor{blue!15}0.20& \cellcolor{blue!15}0.18&
		\cellcolor{blue!15}0.18\\
		$\bm{m}_i \in \mathbb{R}^{16}$			&\cellcolor{blue!15}3000& \cellcolor{blue!15}II
		&\cellcolor{blue!15}Out-of-control 
		& \cellcolor{blue!15}$CDR$
		&\cellcolor{blue!15}0.62
		& \cellcolor{blue!15}/
		&\cellcolor{blue!15}\underline{0.64}
		& \cellcolor{blue!15}0.59
		& \cellcolor{blue!15}0.59
		&\cellcolor{blue!15}0.59\\
		\hhline{~----------}
		&\cellcolor{green!20}4000& \cellcolor{green!20}I&\cellcolor{green!20}In-control&\cellcolor{green!20}$FAR$
		&\cellcolor{green!20}0.05
		& \cellcolor{green!20}/
		&\cellcolor{green!20}\underline{0.05}
		& \cellcolor{green!20}0.05
		&\cellcolor{green!20}0.05
		&\cellcolor{green!20}0.05\\ 
		&\cellcolor{blue!15}4000& \cellcolor{blue!15}II& \cellcolor{blue!15}In-control& \cellcolor{blue!15}$SR$
		 &\cellcolor{blue!15}0.21
		& \cellcolor{blue!15}/
		& \cellcolor{blue!15}\underline{0.23}& \cellcolor{blue!15}0.20& \cellcolor{blue!15}0.18&
		\cellcolor{blue!15}0.18\\
		&\cellcolor{blue!15}4000& \cellcolor{blue!15}II 
		&\cellcolor{blue!15}Out-of-control
		& \cellcolor{blue!15}$CDR$
		&\cellcolor{blue!15}0.62
		& \cellcolor{blue!15}/
		&\cellcolor{blue!15}\underline{0.65}
		& \cellcolor{blue!15}0.60
		& \cellcolor{blue!15}0.60
		&\cellcolor{blue!15}0.59\\
		\hline
	\end{tabular}
	\caption[Monte Carlo study: Performance of $r$ control charts ($\alpha = 0.05$) in Experiment 1.]{Monte Carlo study: Performance of $r$ control charts ($\alpha = 0.05$) in Experiment 1 with reference samples $R$ being predicted classes that were randomly constructed. The underlined numbers indicate the suggested method based on the trade-off between $SR$ and $CDR$. We do not compute \textbf{SD} due to the computational complexity being $O(n^{k+1})$.}
	\label{MonteCarlo}
\end{table}

\section{Experiment 2: Multiclass Classification of Questions}
\label{Experiment2}
For the second experiment, we use the Text REtrieval Conference (TREC) dataset which consists of fact-based questions divided into six broad semantic categories\footnote{https://cogcomp.seas.upenn.edu/Data/QA/QC/} (cf. \citealp{voorhees2000overview}). The model was trained with the four classes: ``Numeric values'', ``Description and abstract concepts'', ``Entities'' and ``Human beings''. Examples of such questions can be found in Table \ref{TREC}. The classification task is to assign an incoming question to one of four categories.

The trained neural network contains three hidden layers: a word embedding layer, a Long Short-Term Memory (LSTM) layer, and a fully connected layer which we use as the embedding generator of the size $1\times8$. After that, the output layer returns softmax vector $1 \times 4$, where the maximum value corresponds to the label of the predicted category. It is worth noting that here the word embedding layer is not a part of our monitoring approach but a Natural Language Processing (NLP) technique that enables the model to associate a numerical vector to every word so that the distance between any two vectors is related to the semantic meaning of the encrypted words (cf. \citealp{yin2018dimensionality}). For the interested reader, we recommend referring to publications that offer a comprehensive introduction to neural networks for NLP tasks, for example, \cite{goldberg2016primer} and \cite{nammous2019natural}.   

In total, 700 data points of each category were used as the training data. The achieved accuracy on the test dataset that contains 150 unseen samples for each class is 81.17\% after 25 training epochs. The 60 out-of-control samples were taken from two other semantic categories that were not used for training, namely ``Abbreviations'' and ``Locations''.

\begin{table}
\renewcommand{\arraystretch}{0.8}
	\centering
	\begin{tabular}{ll } 
		\toprule
	Class                       &   Example                                     \\ 
		\midrule
	Numeric values	                    &  What is the size of Argentina?               \\
	Description and abstract concepts	&   What is artificial intelligence?            \\
    Entities		                    & What is the tallest piece on a chessboard?    \\
    Human beings                        &  Who invented basketball?                     \\
		\bottomrule
	\end{tabular}
\caption{Question examples and respective four categories of the TREC dataset used for training the ANN in Experiment 2.}
\label{TREC}
\end{table}

\subsection{ Monitoring Results}

Considering the results of Experiment 2 in Table \ref{PredClassBestSample}, we notice a decreasing $SR$ when the size of the reference sample is increasing. Nevertheless, the $SR$ values remain substantially high for considered control charts. Although the further increase of reference samples might improve monitoring during the in-control state, it negatively affects the detection of anomalous data. As we can observe in the case of \textbf{HD}$_r$, the $CDR$ is reduced by 25\%, changing from $|R| = 400$ to $|R| = 600$. 

Overall, we notice that either a paired control chart or another procedure to decide confidently when the data points are in- our out-of-control is required to improve the performance in Experiment 2. However, to understand what could be the underlying reasons for unsatisfactory results, we inspect the misclassification effect as well as visually compare the in- and out-of-control data.

\subsection{Misclassification and Data Diagnostic}

\begin{figure}
	\begin{center}
		\includegraphics[width=1.00\textwidth,trim= 0cm 0cm 0cm 0cm,clip]{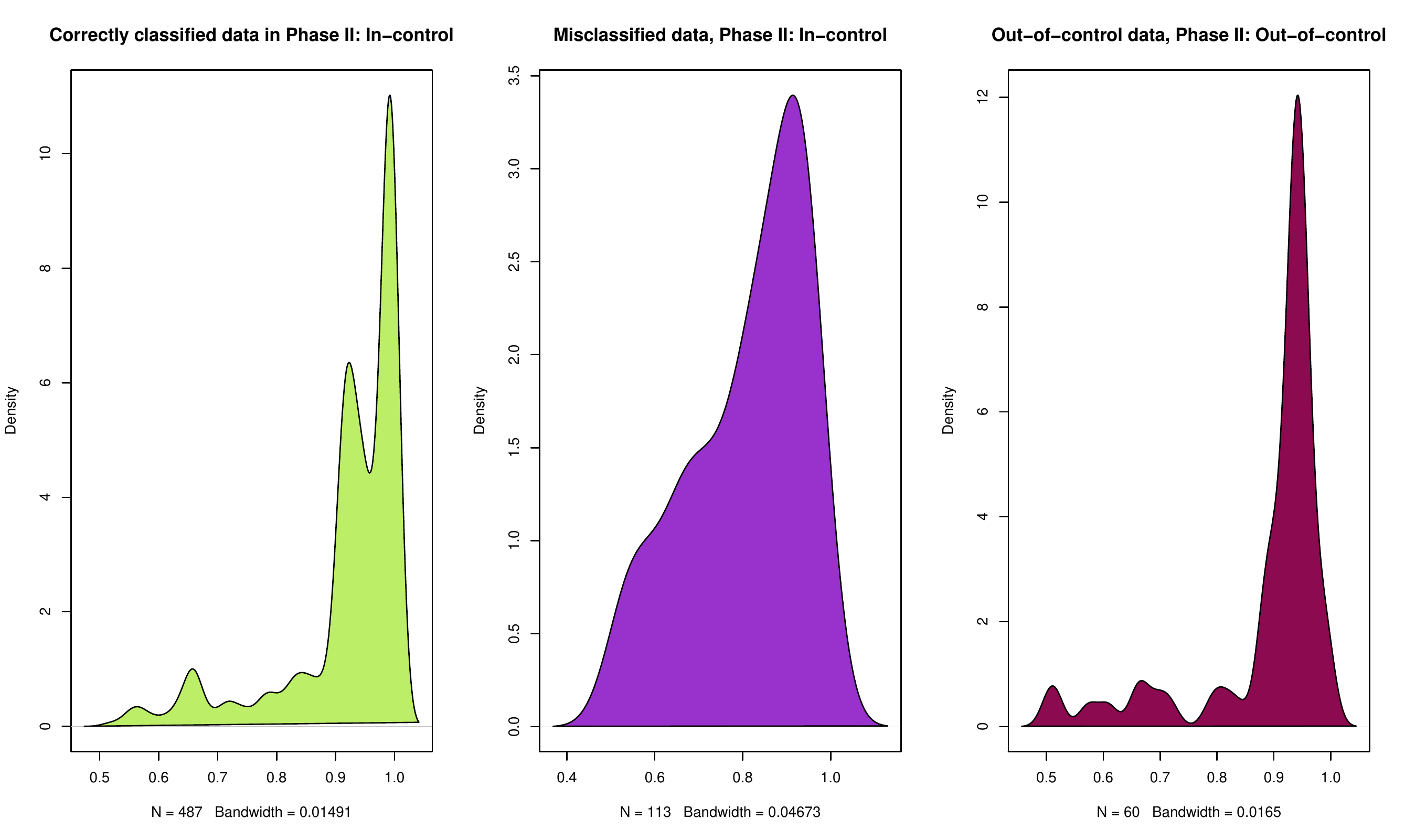} 
		\caption{Kernel density estimates of the score distributions in Phase II for in-control (test data) and out-of-control samples.}
		\label{ScoresDensities}
	\end{center}
\end{figure}

There are 113 samples from the test data that were misclassified, and the softmax output serves as a model's confidence about its prediction in our case. Looking at the density plots in Figure \ref{ScoresDensities}, we notice an evident difference in distributions between correctly classified and misclassified samples in Phase II (in-control). Remarkably, the score of out-of-control samples are rather high and resemble the scores of the in-control samples from Phase II. This behavior can be explained by the similarity of the out-of-control data and the reference samples. More precisely, the network is trained to distinguish the phrases based on features that are identical for both the in-control and out-of-control samples. However, it is worth noting that this could be different on lower aggregation levels (i.e., hidden layers closer to the input).

\begin{figure}
	\begin{center}
		\hspace*{-1.3cm}
		\includegraphics[width=0.80\textwidth, trim= 0cm 0.5cm 0cm 0.5cm,clip]{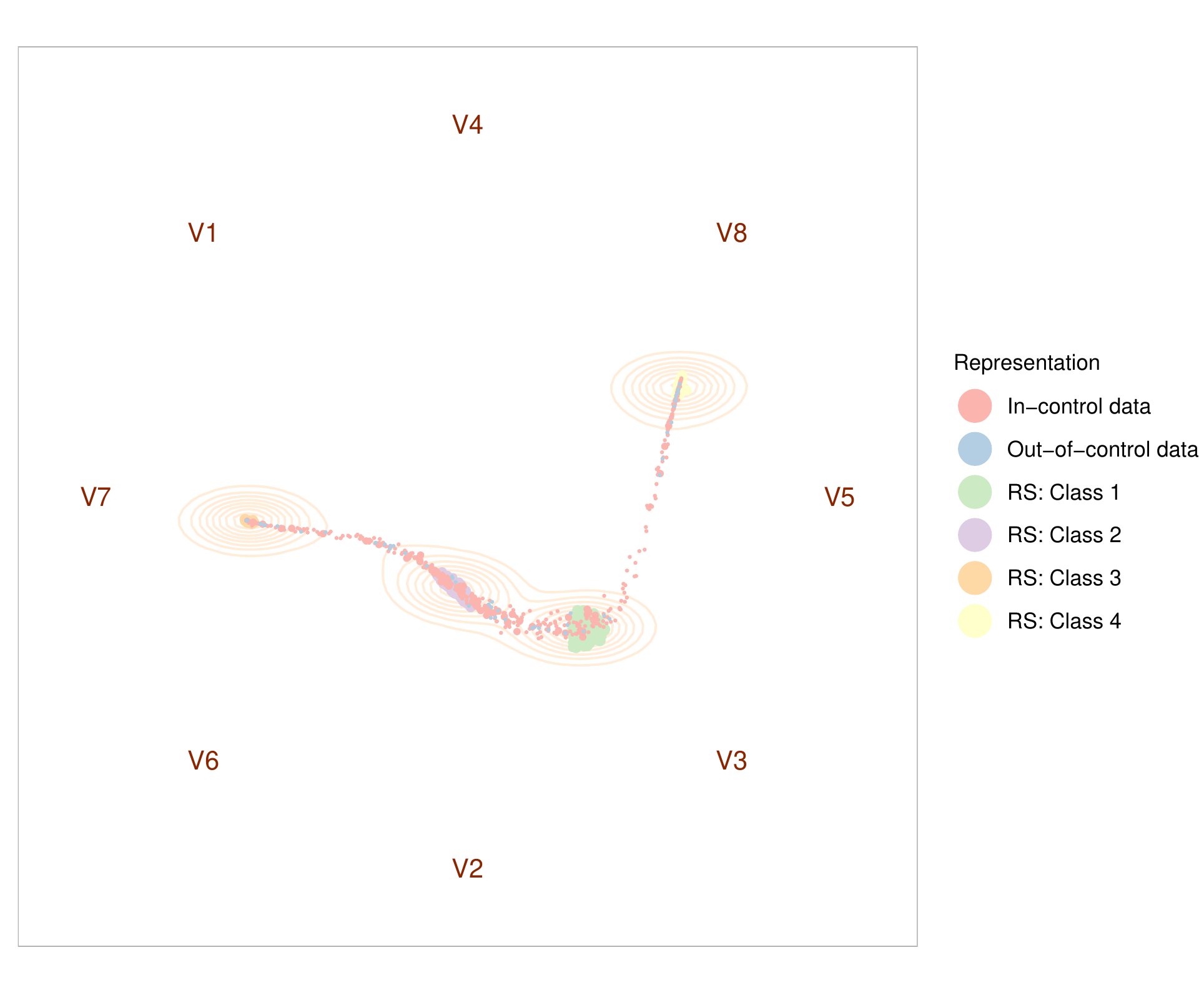}
		\caption{Visualization of the Reference Samples ($RS$) $\{R_1, \dots, R_4\}$ with $|R| = 400$ and the data from Phase II, in-control part (test data) from Experiment 2. $V_1, \dots, V_8$ define anchors which correspond to neurons that produced embeddings $\bm{m}_i \in \mathbb{R}^{8}$. Density contour plots outline respective classes. }
		\label{Plot8D}
	\end{center}
\end{figure}

The Radial Coordinate visualization (Radviz) shown in Figure \ref{Plot8D} reveals that this issue is present in our case: The out-of-control data often overlap with the reference samples. Moreover, most of the in-control (test data) regions are not covered by the reference samples. However, to answer the question of whether the high signal rate is partially due to the misclassification samples, we compute conditional sample rates on misclassification ($SR|M$) and correct prediction ($SR|C$).

As shown in Table \ref{MisclassDiagnostic}, the additional information about misclassified samples could considerably improve the $SR$ results (cf. the outcome for $|R| = 600$ of $SR|C$). Moreover, we notice that the majority of the misclassified samples would be flagged as anomalous, with the signal rates declining when the reference sample size grows. Hence, the combination of our monitoring approach with an additional misclassification detection technique could lead to a more reliable nonstationarity detection.

\begin{table}
	\renewcommand{\arraystretch}{0.8}
	\centering
	\scriptsize
	\begin{tabular}{|c|c|c|c|c|cc:ccccccc|}
		
		\hline
		\multicolumn{1}{|c}{\textbf{Evaluation}} &
		\multicolumn{1}{c}{Size $|R|$} &
		\multicolumn{1}{c}{Phase} &
		\multicolumn{1}{c}{Observed process} &
		\multicolumn{1}{c|}{Metric} &
		\multicolumn{1}{c}{\textbf{MD}} &
		\multicolumn{1}{c:}{\textbf{SD}} &
		\multicolumn{1}{c}{\textbf{HD$_r$}}  &
		\multicolumn{1}{c}{\textbf{PD$_1^a$}} &
		\multicolumn{1}{c}{\textbf{PD$_2^a$}} &
		\multicolumn{1}{c}{\textbf{PD$_3^a$}}&
		\multicolumn{1}{c}{\textbf{PD$_1$}} &
		\multicolumn{1}{c}{\textbf{PD$_2$}} &
		\multicolumn{1}{c|}{\textbf{PD$_3$}}\\

		\hline

		&\cellcolor{green!20}400 &\cellcolor{green!20}I& \cellcolor{green!20} In-control&\cellcolor{green!20}$FAR$ &\cellcolor{green!20}0.05
		&\cellcolor{green!20}/
		&\cellcolor{green!20}\underline{0.05}
		& \cellcolor{green!20}0.05
		&\cellcolor{green!20}0.05
		&\cellcolor{green!20}0.05
		& \cellcolor{green!20}0.05
		&\cellcolor{green!20}0.05
		&\cellcolor{green!20}0.05 \\ 
		& \cellcolor{blue!15}400 &\cellcolor{blue!15}II& \cellcolor{blue!15} In-control& \cellcolor{blue!15}$SR$ & \cellcolor{blue!15}0.68 & \cellcolor{blue!15}/& 
		\cellcolor{blue!15}\underline{0.69}& \cellcolor{blue!15}0.64& \cellcolor{blue!15}0.62&
		\cellcolor{blue!15}0.64& \cellcolor{blue!15}0.65& \cellcolor{blue!15}0.65&
		\cellcolor{blue!15}0.66\\
		& \cellcolor{blue!15}400	&\cellcolor{blue!15}II& \cellcolor{blue!15} Out-of-control
		&\cellcolor{blue!15} $CDR$
		& \cellcolor{blue!15}0.58&\cellcolor{blue!15}/
		&\cellcolor{blue!15}\underline{0.67}
		& \cellcolor{blue!15}0.48
		& \cellcolor{blue!15}0.50
		&\cellcolor{blue!15}0.50
		& \cellcolor{blue!15}0.52
		& \cellcolor{blue!15}0.53
		&\cellcolor{blue!15}0.55\\
		\hhline{~-------------}
		&\cellcolor{green!20}500 	&\cellcolor{green!20}I& \cellcolor{green!20} In-control&\cellcolor{green!20}$FAR$ &\cellcolor{green!20}0.05
		& \cellcolor{green!20}/
		&\cellcolor{green!20}\underline{0.05}
		& \cellcolor{green!20}0.05
		&\cellcolor{green!20}0.05
		&\cellcolor{green!20}0.05
		& \cellcolor{green!20}0.05
		&\cellcolor{green!20}0.05
		&\cellcolor{green!20}0.05 \\ 
		Experiment 2 & \cellcolor{blue!15}500	&\cellcolor{blue!15}II& \cellcolor{blue!15} In-control& \cellcolor{blue!15}$SR$  & \cellcolor{blue!15}0.45 & \cellcolor{blue!15}/&
		\cellcolor{blue!15}\underline{0.60}& \cellcolor{blue!15}0.45& \cellcolor{blue!15}0.43&
		\cellcolor{blue!15}0.44& \cellcolor{blue!15}0.44& \cellcolor{blue!15}0.45&
		\cellcolor{blue!15}0.46 \\
		$\bm{m}_i \in \mathbb{R}^{8}$ 	& \cellcolor{blue!15}500			&\cellcolor{blue!15}II& \cellcolor{blue!15} Out-of-control
		&\cellcolor{blue!15} $CDR$
		& \cellcolor{blue!15}0.45&\cellcolor{blue!15}/
		&\cellcolor{blue!15}\underline{0.58}
		& \cellcolor{blue!15}0.43
		& \cellcolor{blue!15}0.37
		&\cellcolor{blue!15}0.37
		& \cellcolor{blue!15}0.40
		& \cellcolor{blue!15}0.38
		&\cellcolor{blue!15}0.45\\
		
		\hhline{~-------------}
		&\cellcolor{green!20}600	&\cellcolor{green!20}I& \cellcolor{green!20} In-control&\cellcolor{green!20}$FAR$  &\cellcolor{green!20}0.05
		& \cellcolor{green!20}/
		&\cellcolor{green!20}\underline{0.05}
		& \cellcolor{green!20}0.05
		&\cellcolor{green!20}0.05
		&\cellcolor{green!20}0.05
		& \cellcolor{green!20}0.05
		&\cellcolor{green!20}0.05
		&\cellcolor{green!20}0.05 \\ 
		& \cellcolor{blue!15}600 &\cellcolor{blue!15}II& \cellcolor{blue!15} In-control& \cellcolor{blue!15}$SR$ & \cellcolor{blue!15}0.35 & \cellcolor{blue!15}/& 
		\cellcolor{blue!15}\underline{0.37} & \cellcolor{blue!15}0.34& \cellcolor{blue!15}0.34&
		\cellcolor{blue!15}0.33 & \cellcolor{blue!15}0.32& \cellcolor{blue!15}0.32&
		\cellcolor{blue!15}0.33 \\
		& \cellcolor{blue!15}600 &\cellcolor{blue!15}II& \cellcolor{blue!15} Out-of-control
		&\cellcolor{blue!15} $CDR$
		& \cellcolor{blue!15}0.30&\cellcolor{blue!15}/
		&\cellcolor{blue!15}\underline{0.42}
		& \cellcolor{blue!15}0.32
		& \cellcolor{blue!15}0.30
		&\cellcolor{blue!15}0.30
		& \cellcolor{blue!15}0.30
		& \cellcolor{blue!15}0.30
		&\cellcolor{blue!15}0.30 \\
		\hline

		&\cellcolor{green!20}50 &\cellcolor{green!20}I& \cellcolor{green!20} In-control&\cellcolor{green!20}$FAR$ &\cellcolor{green!20}0.04
		&\cellcolor{green!20}0.00
		&\cellcolor{green!20}0.04
		& \cellcolor{green!20}0.04
		&\cellcolor{green!20}0.04
		&\cellcolor{green!20}\underline{0.04}
		& \cellcolor{green!20}0.04
		&\cellcolor{green!20}\underline{0.04}
		&\cellcolor{green!20}\underline{0.04}\\ 
		& \cellcolor{blue!15}50 &\cellcolor{blue!15}II& \cellcolor{blue!15} In-control& \cellcolor{blue!15}$SR$ & \cellcolor{blue!15}0.26 & \cellcolor{blue!15}0.00& 
		\cellcolor{blue!15}0.61& \cellcolor{blue!15}0.16& \cellcolor{blue!15}0.16&
		\cellcolor{blue!15}\underline{0.05}& \cellcolor{blue!15}0.00& \cellcolor{blue!15}\underline{0.05}&
		\cellcolor{blue!15}\underline{0.05}\\
		& \cellcolor{blue!15}50	&\cellcolor{blue!15}II& \cellcolor{blue!15} Out-of-control
		&\cellcolor{blue!15} $CDR$
		& \cellcolor{blue!15}1.00 &\cellcolor{blue!15}0.00
		&\cellcolor{blue!15}1.00
		& \cellcolor{blue!15}1.00
		& \cellcolor{blue!15}1.00
		&\cellcolor{blue!15}\underline{1.00}
		& \cellcolor{blue!15}1.00
		& \cellcolor{blue!15}\underline{1.00}
		&\cellcolor{blue!15}\underline{1.00}\\
		\hhline{~-------------}
		&\cellcolor{green!20}60 	&\cellcolor{green!20}I& \cellcolor{green!20} In-control&\cellcolor{green!20}$FAR$ &\cellcolor{green!20}0.05
		& \cellcolor{green!20}0.00
		&\cellcolor{green!20}0.05
		& \cellcolor{green!20}0.05
		& \cellcolor{green!20}0.05
		&\cellcolor{green!20}0.05
		& \cellcolor{green!20}\underline{0.05}
		&\cellcolor{green!20}\underline{0.05}
		&\cellcolor{green!20}\underline{0.05} \\ 
		Experiment 3 & \cellcolor{blue!15}60	&\cellcolor{blue!15}II& \cellcolor{blue!15} In-control& \cellcolor{blue!15}$SR$  & \cellcolor{blue!15}0.16 & \cellcolor{blue!15}0.00&
		\cellcolor{blue!15} 0.50& \cellcolor{blue!15}0.26& \cellcolor{blue!15}0.24&
		\cellcolor{blue!15} 0.16& \cellcolor{blue!15}\underline{0.00}& \cellcolor{blue!15}\underline{0.00}&
		\cellcolor{blue!15}\underline{0.00}\\
		$\bm{m}_i \in \mathbb{R}^{3}$ 	& \cellcolor{blue!15}60			&\cellcolor{blue!15}II& \cellcolor{blue!15} Out-of-control
		&\cellcolor{blue!15} $CDR$
		& \cellcolor{blue!15}1.00 &\cellcolor{blue!15}0.00
		&\cellcolor{blue!15}1.00
		& \cellcolor{blue!15}1.00
		& \cellcolor{blue!15}1.00
		&\cellcolor{blue!15}1.00
		& \cellcolor{blue!15}\underline{1.00}
		& \cellcolor{blue!15}\underline{1.00}
		&\cellcolor{blue!15}\underline{1.00}\\
		
		\hhline{~-------------}
		&\cellcolor{green!20}70	&\cellcolor{green!20}I& \cellcolor{green!20} In-control&\cellcolor{green!20}$FAR$  &\cellcolor{green!20}\underline{0.04}
		&\cellcolor{green!20}0.00
		&\cellcolor{green!20}0.04
		& \cellcolor{green!20}0.04
		&\cellcolor{green!20}\underline{0.04}
		&\cellcolor{green!20}0.04
		& \cellcolor{green!20}0.04
		&\cellcolor{green!20}0.04
		&\cellcolor{green!20}0.04\\ 
		& \cellcolor{blue!15}70 &\cellcolor{blue!15}II& \cellcolor{blue!15} In-control& \cellcolor{blue!15}$SR$ & \cellcolor{blue!15}\underline{0.05} & \cellcolor{blue!15}0.00& 
		\cellcolor{blue!15}0.37& \cellcolor{blue!15}0.00&
		\cellcolor{blue!15}\underline{0.05}& 
		\cellcolor{blue!15}0.03& \cellcolor{blue!15}0.11&\cellcolor{blue!15}0.11&
		\cellcolor{blue!15}0.11\\
		& \cellcolor{blue!15}70 &\cellcolor{blue!15}II& \cellcolor{blue!15} Out-of-control
		&\cellcolor{blue!15} $CDR$
		& \cellcolor{blue!15}\underline{1.00}&\cellcolor{blue!15}0.00
		&\cellcolor{blue!15}1.00
		& \cellcolor{blue!15}1.00
		& \cellcolor{blue!15}\underline{1.00}
		&\cellcolor{blue!15}1.00
		& \cellcolor{blue!15}1.00
		& \cellcolor{blue!15}1.00
		&\cellcolor{blue!15}1.00\\
		\hline

	\end{tabular}
	\caption[Performance of $r$ control charts with $R$ being the predicted class.]{Performance of $r$ control charts ($\alpha = 0.05$) in the presented experiments with $R$ being the predicted class. Green rows highlight Phase I false alarm rates ($FAR$), and violet define Phase II signal and correct detection rates ($SR$ and $CDR$), respectively. The underlined numbers indicate the suggested method based on the trade-off between $SR$ and $CDR$. We compute \textbf{SD} for $\mathbb{R}^{3}$ only, as computational complexity is $O(n^{k+1})$.}
	\label{PredClassBestSample}
\end{table}

\begin{table}
	\centering
	\scriptsize
	\renewcommand{\arraystretch}{0.8}
	\begin{tabular}{|c|c|c|cc:ccccccc|}
		
		\hline
		\multicolumn{1}{|c}{\textbf{Evaluation}} &
		\multicolumn{1}{c}{Size $|R|$} &
		\multicolumn{1}{c|}{Metric} &
		\multicolumn{1}{c}{\textbf{MD}} &
		\multicolumn{1}{c:}{\textbf{SD}} &
		\multicolumn{1}{c}{\textbf{HD$_r$}}  &
		\multicolumn{1}{c}{\textbf{PD$_1^a$}} &
		\multicolumn{1}{c}{\textbf{PD$_2^a$}} &
		\multicolumn{1}{c}{\textbf{PD$_3^a$}}&
		\multicolumn{1}{c}{\textbf{PD$_1$}} &
		\multicolumn{1}{c}{\textbf{PD$_2$}} &
		\multicolumn{1}{c|}{\textbf{PD$_3$}}\\
		\hline

				\rowcolor{orange!30}
		 \multirow{2}{*}{\cellcolor{white}}&{400}&$SR|M$ & 0.86 & / & 0.87 & 0.81 & 0.82 & 0.84 & 0.83 & 0.86 & 0.86 \\ 
		 \rowcolor{cyan!35}
		 \cellcolor{white}Experiment 2&{400}&$SR|C$ & 0.63 & /  & 0.65 & 0.60 & 0.57 & 0.60 & 0.61 & 0.60 & 0.61 \\ 
		 \hhline{~-----------}
		\rowcolor{orange!30}
		\multirow{2}{*}{\cellcolor{white} }&{500}&$SR|M$ & 0.80 & /  & 0.84 & 0.79 & 0.76 & 0.76 & 0.76 & 0.78 & 0.79 \\ 
		 \rowcolor{cyan!35}
		 \cellcolor{white} $\bm{m}_i \in \mathbb{R}^{8}$ &{500}&$SR|C$ & 0.37 & /  & 0.55 & 0.37 & 0.35 & 0.36 & 0.37 & 0.37 & 0.38 \\ 
		\hhline{~-----------}
		\rowcolor{orange!30}
		\multirow{2}{*}{\cellcolor{white}}&{600}&$SR|M$ & 0.69 & / & 0.72 & 0.64 & 0.65 & 0.61 & 0.65 & 0.64 & 0.64 \\ 
		 \rowcolor{cyan!35}
		 \cellcolor{white}&{600}&$SR|C$ & 0.28 & / & 0.29 & 0.26 & 0.26 & 0.26 & 0.25 & 0.25 & 0.26 \\ 
		\hline
	\end{tabular}
	\caption[Summary of signal rates under condition the samples were misclassified ($SR|M$) or correctly classified
	 in Phase II ($SR|C$).]{Summary of signal rates under the condition that the samples were misclassified ($SR|M$) or correctly classified ($SR|C$) in Phase II. We do not compute \textbf{SD} due to the computational complexity being $O(n^{k+1})$.}
	\label{MisclassDiagnostic}
\end{table}

\section{Experiment 3: Binary Classification of Sonar Signals}
\label{Experiment3}
In the third experiment, we consider a binary classification problem of sonar data \citep{Dua:2019}. This dataset summarizes sonar signals collected from metal cylinders and cylindrically shaped rocks \citep{gorman1988analysis}. There are 208 samples in total, comprising 111 metal cylinders and 97 rock returns. Each sample consists of a series of 60 numbers ranging from 0.0 to 1.0, representing a normalized spectral envelope. The task of the classifier is to distinguish which samples are from scanning a rock and which are from a metal cylinder.

Our model is an FNN with the architecture $60\rightarrow30\rightarrow15\rightarrow3\rightarrow1$ that comprises four fully connected layers reducing the complexity $1\times60$ of the input data by first processing it through the hidden layers that have 30 and 15 neurons. Afterward, the compressed data representation enters the layer with 3 neurons whose output is also used as embeddings of size $1\times3$ for the monitoring procedure. Then, to obtain the class label, we transform the interim output from size $1\times3$ to $1\times1$. As we have a binary classification problem, we use only one neuron in the output layer together with the sigmoid activation function that is centered around 0.5, returning the probability of the processed sample belonging to class 2. Thus, if the result of the output layer is 0.5 or higher, we conclude that the processed sample is a part of class 2 (rock) and of class 1 (metal cylinder) otherwise. Due to the small size of the dataset, only 38 samples (24 of class 1 and 14 of class 2) are allocated to the testing stage which is later used in Phase II as the in-control data. Consequently, the remaining 85 metal cylinders and 85 rock examples are taken for training the FNN. The performance metrics such as validation loss and accuracy are used to determine the number of epochs, i.e., training cycles in which the model learns from the data and updates the parameters. Following that, the FNN model was trained for 39 epochs and achieved 81.58\% accuracy on the test data. 

To create out-of-control samples, by flattening the input we estimate the parameters of a beta distribution for each class (i.e., $\tilde{\alpha}_1$, $\beta_1$ of class 1, and $\tilde{\alpha}_2$, $\beta_2$ of class 2) and randomly sample from a beta distribution with parameters $\tilde{\alpha}_\nu = \tilde{\alpha}_1/\tilde{\alpha}_2$ and $\beta_\nu = \beta_1/\beta_2$ to generate 30 out of control observations. 

\begin{figure}
	\begin{center}
		\includegraphics[width=1.00\textwidth]{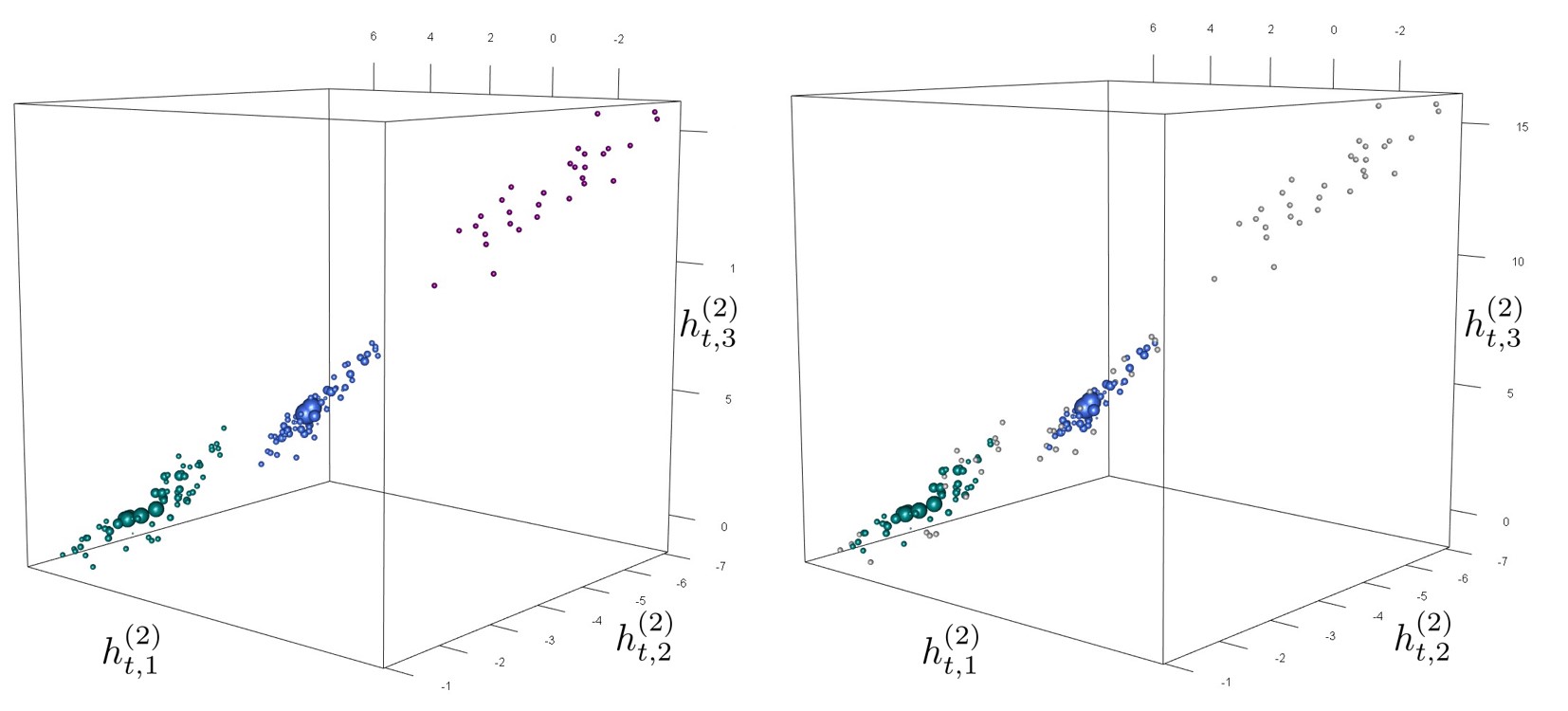} 
		\caption{Reference samples ($|R| = 70$) and out-of-control embeddings from Experiment 3. Blue-colored points belong to class 1, green-colored to class 2, and magenta-colored are out-of-control samples. The size of the points is proportional to the value of Simplicial depth and the gray-colored points highlight the data samples that received $\textbf{SD}(\bm{m}_t)=0$.}
		\label{Exp3SD}
	\end{center}
\end{figure}

Studying the outcomes of Phase I in Experiment 3, Table \ref{PredClassBestSample} we notice that although we choose $\alpha = 0.05$, $FAR$ sometimes equals 0.04. This happens due to rounding, thus, $FAR = 0.04$ for $|R| = 50$ and $|R| = 70$ coincides with the expected value. Consequently, the expected value of $FAR$ is reached for each notion of data depth except Simplicial depth. The reason for that is the large number of data points in the reference sample with $\textbf{SD}(\bm{m}_t)=0$. That can be explained by Simplicial depth assigning zero to every point in the space that lies outside the convex hull of the sample \citep{afshani2016approximating,francisci2019generalization}. In Figure \ref{Exp3SD}, we observe the dispersion of the data and how many samples obtained $\textbf{SD}(\bm{m}_t)=0$. All out-of-control samples received the predictions of class 1, meaning that their depth is determined with respect to the blue point cloud. Overall, the control charts operate well with $|R| = 50$ and the symmetric projection depth, especially with \textbf{PD}$_2$ and \textbf{PD}$_3$. Alternatively, one can choose \textbf{MD} with $|R| = 70$ because of correctly reached $SR$ and similarly high $CDR$.

\section{The $Q$ Control Chart (Batch Size $> 1$)}
Similarly to the $r$ control chart, the $Q$ control chart proposed by \cite{liu1995control} is based on ranks of multivariate observations which are obtained by computing data depth. The test statistic of the $Q$ control chart is the average of consecutive subsets of $r_{\cdot}^c(\bm{m}_i)$, being
\begin{equation*}
Q_{\cdot}^c(\bm{m}_i) = \dfrac{1}{n}\sum_{j=1}^{n}r_{\cdot}^c(\bm{m}_{ij})
\end{equation*}
with the batch size $n$ \citep{liu1995control}. The interpretation of the ranks in the $Q$ control chart is similar to the interpretation in the case of the $r$ control chart.

To compute the control limit, we use the equation
\begin{equation*}
	LCL = \dfrac{(n!\alpha)^{1/n}}{n},
\end{equation*}
given that $\alpha \leq \frac{1}{n!}$ \citep{stoumbos2001nonparametric}. However, if $\alpha > \frac{1}{n!}$, the LCL has to be computed numerically by solving the polynomial equation provided by \cite{liu1995control}. In general,  the process is considered to be out of control if $Q_{\cdot}^c(\bm{m}_i) \leq LCL$.  

To evaluate the performance of $Q$ control charts, we chose those $|R|$ which achieved the most satisfactory (trade-off) performance with $r$ control charts. Due to the small size of the dataset in Experiment 3, we compute the $Q$ control charts with the batch size $n = 5$ only for Experiments 1 and 2.

Considering the results in Table \ref{QChart}, we notice that (apart from \textbf{SD}) the $SR$ values are excessively high. The increase can be explained by a substantial change in control limits. Referring to the previous description, for $n = 3$, we obtain $LCL = 0.22$, and for $n = 5$, $LCL = 0.29$. At the same time, in all possible consolidations, the $Q$ control chart achieves notable results during the out-of-control period, considerably improving the $CDR$ values in Experiment 2 for Projection depth. Thus, if a supplementary procedure can be developed for detecting and filtering signals successfully when the process remains in control, the $Q$ control chart would outperform the $r$ control chart.

\begin{table}
	\centering
	\scriptsize
	\renewcommand{\arraystretch}{0.8}
	\begin{tabular}{|c|c|c|c|c|cc:ccccccc|}
		
		\hline
		\multicolumn{1}{|c}{\textbf{Evaluation}} &
		\multicolumn{1}{c}{Batch size} &
		\multicolumn{1}{c}{Phase} &
		\multicolumn{1}{c}{Observed process} &
		\multicolumn{1}{c|}{Metric} &
		\multicolumn{1}{c}{\textbf{MD}} &
		\multicolumn{1}{c:}{\textbf{SD}} &
		\multicolumn{1}{c}{\textbf{HD$_r$}}  &
			\multicolumn{1}{c}{\textbf{PD$_1^a$}} &
		\multicolumn{1}{c}{\textbf{PD$_2^a$}} &
		\multicolumn{1}{c}{\textbf{PD$_3^a$}} &
		\multicolumn{1}{c}{\textbf{PD$_1$}} &
		\multicolumn{1}{c}{\textbf{PD$_2$}} &
		\multicolumn{1}{c|}{\textbf{PD$_3$}}\\

		\hline

		&\cellcolor{green!20}3	&\cellcolor{green!20}I& \cellcolor{green!20} In-control&\cellcolor{green!20}$FAR$ &\cellcolor{green!20}0.08
		& \cellcolor{green!20}/
		&\cellcolor{green!20}0.09
		& \cellcolor{green!20}\underline{0.07}
		&\cellcolor{green!20}0.08
		&\cellcolor{green!20}0.08
		& \cellcolor{green!20}0.08
		&\cellcolor{green!20}0.08
		&\cellcolor{green!20}0.08 \\ 
		Experiment 1 & \cellcolor{blue!15}3	&\cellcolor{blue!15}II& \cellcolor{blue!15} In-control& \cellcolor{blue!15}$SR$  & \cellcolor{blue!15}0.58& \cellcolor{blue!15}/&
		\cellcolor{blue!15}0.59& \cellcolor{blue!15}\underline{0.55}&\cellcolor{blue!15}0.56&\cellcolor{blue!15}0.56&
		\cellcolor{blue!15}0.55& \cellcolor{blue!15}0.54&
		\cellcolor{blue!15}0.55\\
		$\bm{m}_i \in \mathbb{R}^{16}$ 	& \cellcolor{blue!15}3		&\cellcolor{blue!15}II& \cellcolor{blue!15} Out-of-control
		&\cellcolor{blue!15} $CDR$
		& \cellcolor{blue!15}0.98&\cellcolor{blue!15}/
		&\cellcolor{blue!15}0.98
		& \cellcolor{blue!15}\underline{0.98}
		& \cellcolor{blue!15}0.97
		& \cellcolor{blue!15}0.97
		& \cellcolor{blue!15}0.98
		& \cellcolor{blue!15}0.96
		&\cellcolor{blue!15}0.97\\
		
		\hhline{~-------------}
		$|R| = 3000$		&\cellcolor{green!20}5	&\cellcolor{green!20}I& \cellcolor{green!20} In-control&\cellcolor{green!20}$FAR$  &\cellcolor{green!20}0.10
		& \cellcolor{green!20}/
		&\cellcolor{green!20}0.13
		& \cellcolor{green!20}0.10
		&\cellcolor{green!20}0.10
		&\cellcolor{green!20}0.11
		& \cellcolor{green!20}\underline{0.10}
		&\cellcolor{green!20}\underline{0.10}
		&\cellcolor{green!20}0.11 \\ 
		& \cellcolor{blue!15}5 &\cellcolor{blue!15}II& \cellcolor{blue!15} In-control& \cellcolor{blue!15}$SR$ & \cellcolor{blue!15}0.75& \cellcolor{blue!15}/& 
		\cellcolor{blue!15}0.76&  
		\cellcolor{blue!15}0.72&\cellcolor{blue!15}0.73&\cellcolor{blue!15}0.73&\cellcolor{blue!15}\underline{0.71}& \cellcolor{blue!15}\underline{0.71}&
		\cellcolor{blue!15}0.71\\
		& \cellcolor{blue!15}5&\cellcolor{blue!15}II& \cellcolor{blue!15} Out-of-control
		&\cellcolor{blue!15} $CDR$
		& \cellcolor{blue!15}1.00&\cellcolor{blue!15}/
		&\cellcolor{blue!15}1.00
		& \cellcolor{blue!15}1.00
		& \cellcolor{blue!15}1.00
		&\cellcolor{blue!15}1.00
		& \cellcolor{blue!15}\underline{1.00}
		& \cellcolor{blue!15}\underline{1.00}
		&\cellcolor{blue!15}1.00\\
		\hline

		&\cellcolor{green!20}3	&\cellcolor{green!20}I& \cellcolor{green!20} In-control&\cellcolor{green!20}$FAR$ &\cellcolor{green!20}0.15
		& \cellcolor{green!20}/
		&\cellcolor{green!20}\underline{0.16}
		& \cellcolor{green!20}0.13
		&\cellcolor{green!20}0.11
		&\cellcolor{green!20}0.11 
		& \cellcolor{green!20}0.15
		&\cellcolor{green!20}0.14
		&\cellcolor{green!20}0.14\\ 
		Experiment 2 & \cellcolor{blue!15}3	&\cellcolor{blue!15}II& \cellcolor{blue!15} In-control& \cellcolor{blue!15}$SR$  & \cellcolor{blue!15}0.73& \cellcolor{blue!15}/&
		\cellcolor{blue!15}\underline{0.75}& \cellcolor{blue!15}0.72& \cellcolor{blue!15}0.71&
		\cellcolor{blue!15}0.72& \cellcolor{blue!15}0.73& \cellcolor{blue!15}0.72&
		\cellcolor{blue!15}0.72\\
		$\bm{m}_i \in \mathbb{R}^{8}$ 	& \cellcolor{blue!15}3		&\cellcolor{blue!15}II& \cellcolor{blue!15} Out-of-control
		&\cellcolor{blue!15} $CDR$
		& \cellcolor{blue!15}0.67&\cellcolor{blue!15}/
		&\cellcolor{blue!15}\underline{0.72}
		& \cellcolor{blue!15}0.62
		& \cellcolor{blue!15}0.57
		&\cellcolor{blue!15}0.62
		& \cellcolor{blue!15}0.67
		& \cellcolor{blue!15}0.57
		&\cellcolor{blue!15}0.72\\
		
		\hhline{~-------------}
		$|R| = 400$		&\cellcolor{green!20}5	&\cellcolor{green!20}I& \cellcolor{green!20} In-control&\cellcolor{green!20}$FAR$  &\cellcolor{green!20}0.19
		& \cellcolor{green!20}/
		&\cellcolor{green!20}\underline{0.19}
		& \cellcolor{green!20}0.17
		&\cellcolor{green!20}0.15
		&\cellcolor{green!20}0.16
		& \cellcolor{green!20}0.20
		&\cellcolor{green!20}0.18
		&\cellcolor{green!20}0.18 \\ 
		& \cellcolor{blue!15}5 &\cellcolor{blue!15}II& \cellcolor{blue!15} In-control& \cellcolor{blue!15}$SR$ & \cellcolor{blue!15}0.74& \cellcolor{blue!15}/& 
		\cellcolor{blue!15}\underline{0.77}& \cellcolor{blue!15}0.74& \cellcolor{blue!15}0.74&
		\cellcolor{blue!15}0.74& \cellcolor{blue!15}0.74& \cellcolor{blue!15}0.75&
		\cellcolor{blue!15}0.74\\
		& \cellcolor{blue!15}5&\cellcolor{blue!15}II& \cellcolor{blue!15} Out-of-control
		&\cellcolor{blue!15} $CDR$
		& \cellcolor{blue!15}0.80 &\cellcolor{blue!15}/
		&\cellcolor{blue!15}\underline{0.90}
		& \cellcolor{blue!15}0.80
		& \cellcolor{blue!15}0.70
		&\cellcolor{blue!15}0.80 
		& \cellcolor{blue!15}0.80
		& \cellcolor{blue!15}0.80
		&\cellcolor{blue!15}0.80 \\
		\hline	
		Experiment 3		&\cellcolor{green!20}3 &\cellcolor{green!20}I& \cellcolor{green!20} In-control&\cellcolor{green!20}$FAR$ &\cellcolor{green!20}0.07
		&\cellcolor{green!20}0.00
		&\cellcolor{green!20}0.11
		& \cellcolor{green!20}\underline{0.02}
		&\cellcolor{green!20}\underline{0.02}
		&\cellcolor{green!20}\underline{0.02}
		& \cellcolor{green!20}0.04
		&\cellcolor{green!20}0.04
		&\cellcolor{green!20}0.04 \\ 
		$\bm{m}_i \in \mathbb{R}^{3}$		& \cellcolor{blue!15}3 &\cellcolor{blue!15}II& \cellcolor{blue!15} In-control& \cellcolor{blue!15}$SR$ & \cellcolor{blue!15}0.36& \cellcolor{blue!15}0.00& 
		\cellcolor{blue!15}0.34& \cellcolor{blue!15}\underline{0.18}& \cellcolor{blue!15}\underline{0.18}&
			\cellcolor{blue!15}\underline{0.18}& \cellcolor{blue!15}0.36& \cellcolor{blue!15}0.36&
		\cellcolor{blue!15}0.27\\
		$|R| = 70$ 		& \cellcolor{blue!15}3	&\cellcolor{blue!15}II& \cellcolor{blue!15} Out-of-control
		&\cellcolor{blue!15} $CDR$
		& \cellcolor{blue!15}1.00&\cellcolor{blue!15}0.00
		&\cellcolor{blue!15}1.00
		& \cellcolor{blue!15}\underline{1.00}
		& \cellcolor{blue!15}\underline{1.00}
		&\cellcolor{blue!15}\underline{1.00}
		& \cellcolor{blue!15}1.00
		& \cellcolor{blue!15}1.00
		&\cellcolor{blue!15}1.00\\
	
		\hline
	\end{tabular}
	\caption[Performance of $Q$ control charts.]{Performance of $Q$ control charts ($LCL = 0.22$ for $n = 3$ and $LCL = 0.29$ for $n = 5$) in the presented experiments with $R$ being the predicted class. The underlined numbers indicate the suggested method based on the trade-off between $SR$ and $CDR$. We compute \textbf{SD} for $\mathbb{R}^{3}$ only, as computational complexity is $O(n^{k+1})$.}
	\label{QChart}
\end{table}

\section{Reference Sample in Form of Merged Classes}
In this analytical part, we monitor the data in each of the three experiments by creating the reference samples without conditioning on the (predicted) class $c$. That is, the Phase II samples are compared to a joint embedding distribution from Phase I, being independent of the class labels. The benefit of this approach is that no predictions are needed, meaning that if the ANN model provides an incorrect class label, the possible negative effect of misclassification is excluded. Additionally, the application of merged reference samples implies that if a data point is flagged as out of control, it would remain out of control compared to the entire reference data. 

Reporting one case for each experiment in Table \ref{MergedCaseTable}, we can observe satisfactory performance in Experiment 3. In Experiments 1 and 2, the results are less convincing during the out-of-control part. The reason is that the data depth values of reference sample points in a merged case are considerably lower than in a case of individual classes, leading to a less sensitive detection of nonstationary samples. On the contrary, the $SR$ values are reduced compared to the case when the reference sample relates to the predicted class only.

Although the calculation of the data depth with respect to a merged version of a
reference sample eliminates the misclassification problem, in our empirical study the
detection results of spurious data by using predicted classes on their own are
substantially better for high-dimensional problems. For low-dimensional cases such as Experiment 3, we recommend first examining the performance of the monitoring based on the merged reference sample of different sizes, and then, if it is not operating acceptably, applying the method with the reference samples of predicted classes.

\begin{table}
	\centering
	\scriptsize
	\renewcommand{\arraystretch}{0.8}
	\begin{tabular}{|c|c|c|c|c|cc:ccccccc|}
		
		\hline
		\multicolumn{1}{|c}{\textbf{Evaluation}} &
		\multicolumn{1}{c}{Size $|R|$} &
		\multicolumn{1}{c}{Phase} &
		\multicolumn{1}{c}{Observed process} &
		\multicolumn{1}{c|}{Metric} &
		\multicolumn{1}{c}{\textbf{MD}} &
		\multicolumn{1}{c:}{\textbf{SD}} &
		\multicolumn{1}{c}{\textbf{HD$_r$}}  &
		\multicolumn{1}{c}{\textbf{PD$_1^a$}} &
		\multicolumn{1}{c}{\textbf{PD$_2^a$}} &
		\multicolumn{1}{c}{\textbf{PD$_3^a$}}&
		\multicolumn{1}{c}{\textbf{PD$_1$}} &
		\multicolumn{1}{c}{\textbf{PD$_2$}} &
		\multicolumn{1}{c|}{\textbf{PD$_3$}}\\

		\hline

			&\cellcolor{green!20}30000 	&\cellcolor{green!20}I& \cellcolor{green!20} In-control&\cellcolor{green!20}$FAR$ &\cellcolor{green!20}0.05
		& \cellcolor{green!20}/
		&\cellcolor{green!20}\underline{0.05}
		& \cellcolor{green!20}0.05
		&\cellcolor{green!20}0.05
		&\cellcolor{green!20}0.05 
		& \cellcolor{green!20}0.05
		&\cellcolor{green!20}0.05
		&\cellcolor{green!20}0.05\\ 
		Experiment 1& \cellcolor{blue!15}30000	&\cellcolor{blue!15}II& \cellcolor{blue!15} In-control& \cellcolor{blue!15}$SR$  & \cellcolor{blue!15}0.15& \cellcolor{blue!15}/&
		\cellcolor{blue!15}\underline{0.20}&
		\cellcolor{blue!15}0.10& \cellcolor{blue!15}0.09&
		\cellcolor{blue!15}0.08&
		\cellcolor{blue!15}0.09& \cellcolor{blue!15}0.09&
		\cellcolor{blue!15}0.09\\
		$\bm{m}_i \in \mathbb{R}^{16}$ 	& \cellcolor{blue!15}30000			&\cellcolor{blue!15}II& \cellcolor{blue!15} Out-of-control
		&\cellcolor{blue!15} $CDR$
		& \cellcolor{blue!15}0.19&\cellcolor{blue!15}/
		&\cellcolor{blue!15}\underline{0.24}
		& \cellcolor{blue!15}0.02
		& \cellcolor{blue!15}0.02
		&\cellcolor{blue!15}0.01
		& \cellcolor{blue!15}0.02
		& \cellcolor{blue!15}0.02
		&\cellcolor{blue!15}0.02\\
		\hline

		&\cellcolor{green!20}1600 &\cellcolor{green!20}I& \cellcolor{green!20} In-control&\cellcolor{green!20}$FAR$ &\cellcolor{green!20}\underline{0.05}
		&\cellcolor{green!20}/
		&\cellcolor{green!20}0.05
		& \cellcolor{green!20}0.05
		&\cellcolor{green!20}0.05
		&\cellcolor{green!20}0.05 
		& \cellcolor{green!20}0.05
		&\cellcolor{green!20}0.05
		&\cellcolor{green!20}0.05 \\ 
		Experiment 2& \cellcolor{blue!15}1600 &\cellcolor{blue!15}II& \cellcolor{blue!15} In-control& \cellcolor{blue!15}$SR$ & \cellcolor{blue!15}\underline{0.66}& \cellcolor{blue!15}/& 
		\cellcolor{blue!15}0.08&
		\cellcolor{blue!15}0.02
		&\cellcolor{blue!15}0.00
		&\cellcolor{blue!15}0.00
		&\cellcolor{blue!15}0.02& \cellcolor{blue!15}0.08&
		\cellcolor{blue!15}0.00\\ 
		$\bm{m}_i \in \mathbb{R}^{8}$ 		& \cellcolor{blue!15}1600	&\cellcolor{blue!15}II& \cellcolor{blue!15} Out-of-control
		&\cellcolor{blue!15} $CDR$
		& \cellcolor{blue!15}\underline{0.63}&\cellcolor{blue!15}/
		&\cellcolor{blue!15}0.05
		& \cellcolor{blue!15}0.00
		&\cellcolor{blue!15}0.00
		&\cellcolor{blue!15}0.00
		& \cellcolor{blue!15}0.02
		& \cellcolor{blue!15}0.02
		&\cellcolor{blue!15}0.02 \\

		\hline
		
			&\cellcolor{green!20}140	&\cellcolor{green!20}I& \cellcolor{green!20} In-control&\cellcolor{green!20}$FAR$  &\cellcolor{green!20}0.05
		& \cellcolor{green!20}0.00
		&\cellcolor{green!20}0.05
		& \cellcolor{green!20}0.05
		&\cellcolor{green!20}\underline{0.05}
		&\cellcolor{green!20}\underline{0.05} 
		& \cellcolor{green!20}0.05
		&\cellcolor{green!20}0.05
		&\cellcolor{green!20}0.05 \\ 
		Experiment 3& \cellcolor{blue!15}140 &\cellcolor{blue!15}II& \cellcolor{blue!15} In-control& \cellcolor{blue!15}$SR$ & \cellcolor{blue!15}0.08 & \cellcolor{blue!15}0.00& 
		\cellcolor{blue!15}0.00& 
		\cellcolor{blue!15}0.03& 
		\cellcolor{blue!15}\underline{0.05}&
		\cellcolor{blue!15}\underline{0.05}&
		\cellcolor{blue!15}0.08& 
		\cellcolor{blue!15}0.08&
		\cellcolor{blue!15}0.08\\
		$\bm{m}_i \in \mathbb{R}^{3}$ 			& \cellcolor{blue!15}140 &\cellcolor{blue!15}II& \cellcolor{blue!15} Out-of-control
		&\cellcolor{blue!15} $CDR$
		& \cellcolor{blue!15}1.00 &\cellcolor{blue!15}0.00
		&\cellcolor{blue!15}1.00
		& \cellcolor{blue!15}1.00
		& \cellcolor{blue!15}\underline{1.00}
		&\cellcolor{blue!15}\underline{1.00}
		& \cellcolor{blue!15}1.00
		& \cellcolor{blue!15}1.00
		&\cellcolor{blue!15}1.00\\

		\hline

	\end{tabular}
	\caption[Performance of $r$ control charts ($\alpha = 0.05$) in the presented experiments with $R$ being merged classes.]{Performance of $r$ control charts ($\alpha = 0.05$) in the presented experiments with $R$ being merged classes. The underlined numbers indicate the suggested method based on the trade-off between $SR$ and $CDR$. We compute \textbf{SD} for $\mathbb{R}^{3}$ only, as computational complexity is $O(n^{k+1})$.}
	\label{MergedCaseTable}
\end{table}

\section{Computation Time}
\label{SupplMaterial:Time}

For performing online surveillance, the computation time of the monitoring statistic is of particular importance. As the most time-consuming part of our approach is the derivation of data depth values, we compare the execution time of different algorithms to obtain the depth of one data point. To provide a concise summary, we compare running times for the middle sizes of reference samples, namely $|R| = \{3000, 500, 60\}$, and for the cases displayed in Table \ref{MergedCaseTable}. In the case of Projection depth, we present the results of the symmetric type with three applied algorithms. 

\begin{figure}%
    \centering
    \subfloat[\centering Single class reference samples of size $|R| = 3000$]{{\includegraphics[width=0.49\textwidth]{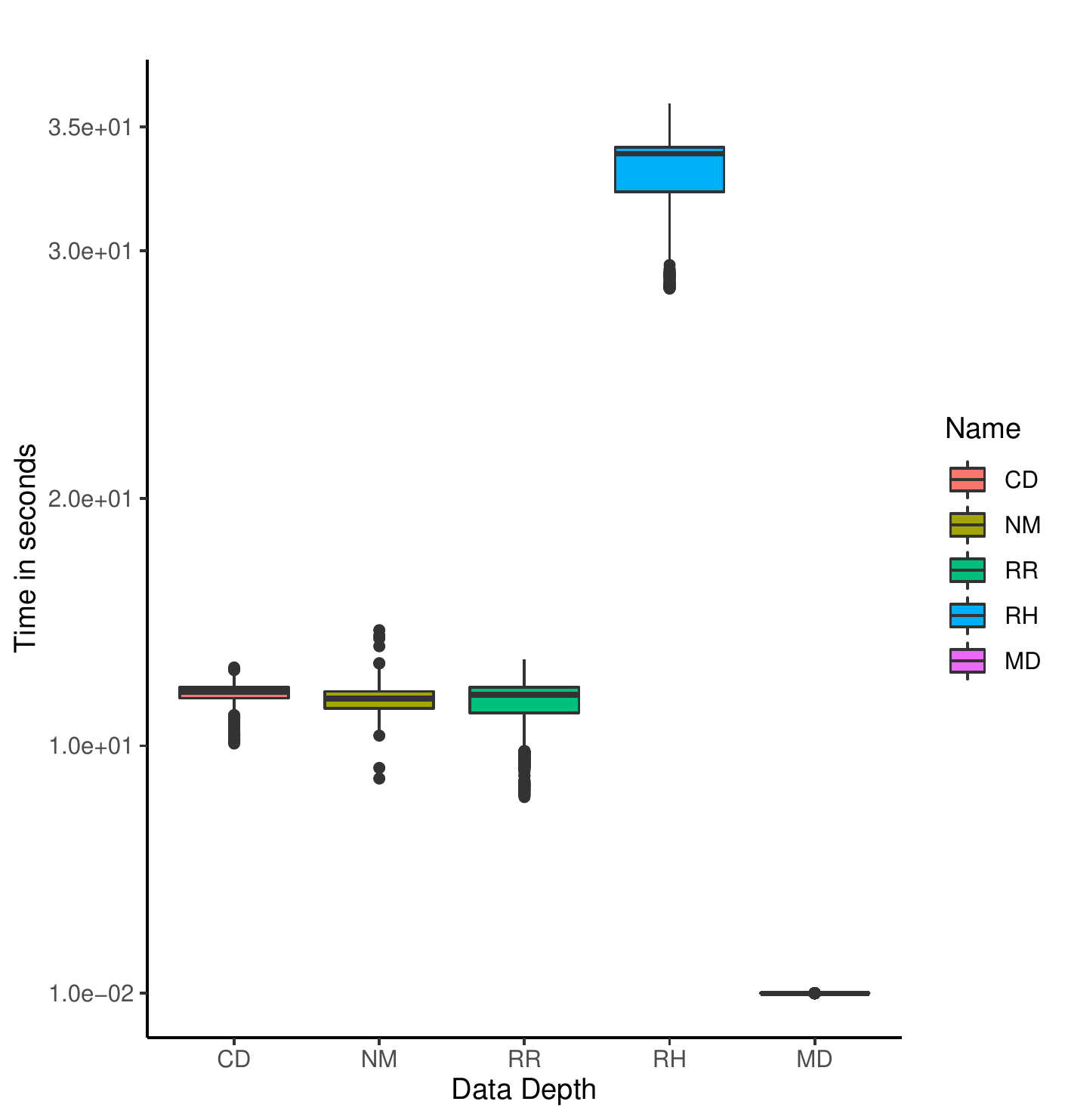} }}%
    \subfloat[\centering Merged reference samples of size $|R| = 30000$]{{\includegraphics[width=0.49\textwidth]{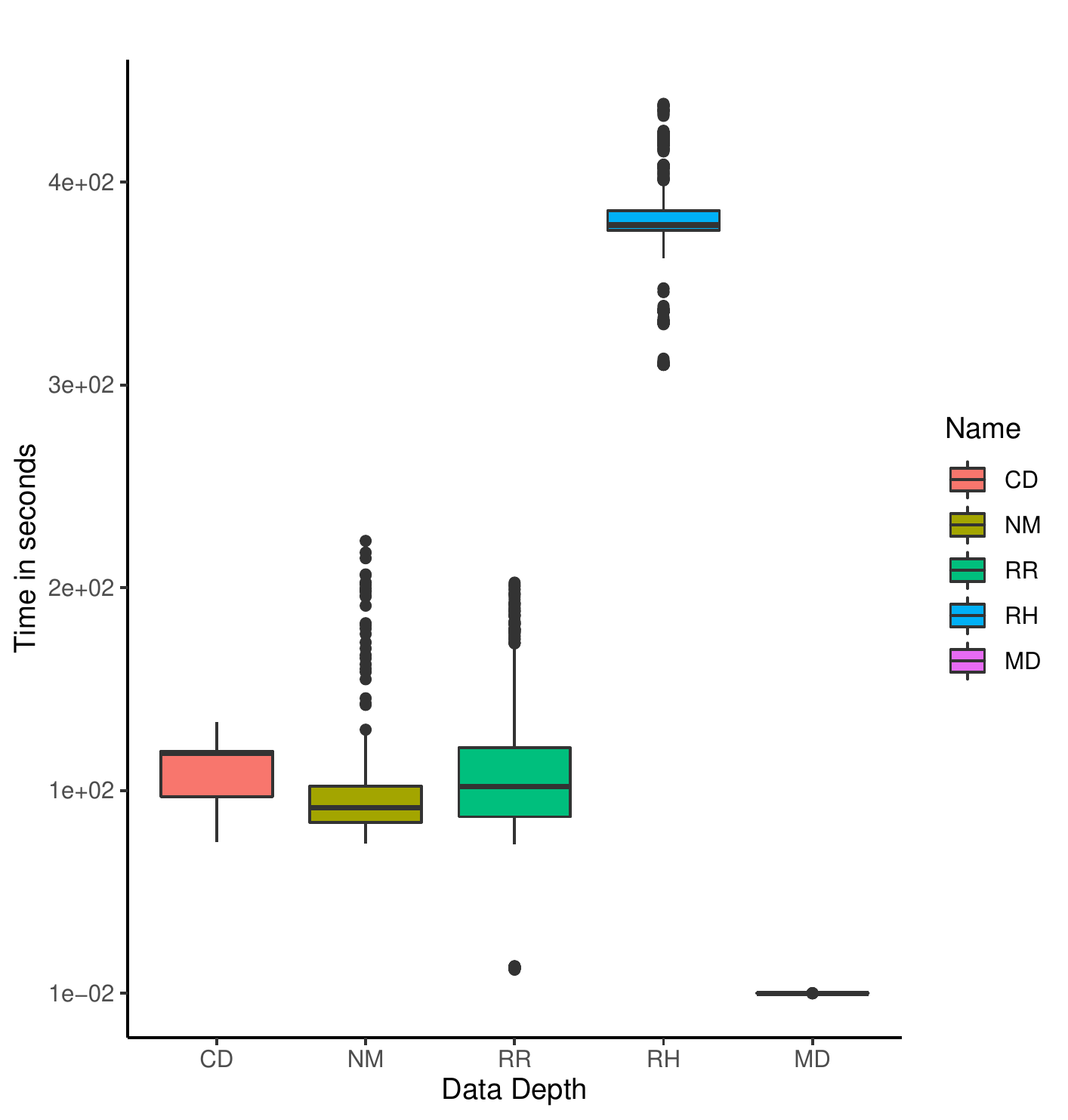} }}%
    \caption{Distribution of computation time for different data depths in Experiment 1. The order is symmetric Projection Depth with \textbf{C}oordinate \textbf{D}escent, \textbf{N}elder-\textbf{M}ead and \textbf{R}efined \textbf{R}andom  algorithms, \textbf{R}obust \textbf{H}alfspace and \textbf{M}ahalanobis \textbf{D}epths.}%
    \label{Exp1Time}%
\end{figure}

\begin{figure}%
    \centering
    \subfloat[\centering Single class reference samples of size $|R| = 500$]{{\includegraphics[width=0.49\textwidth]{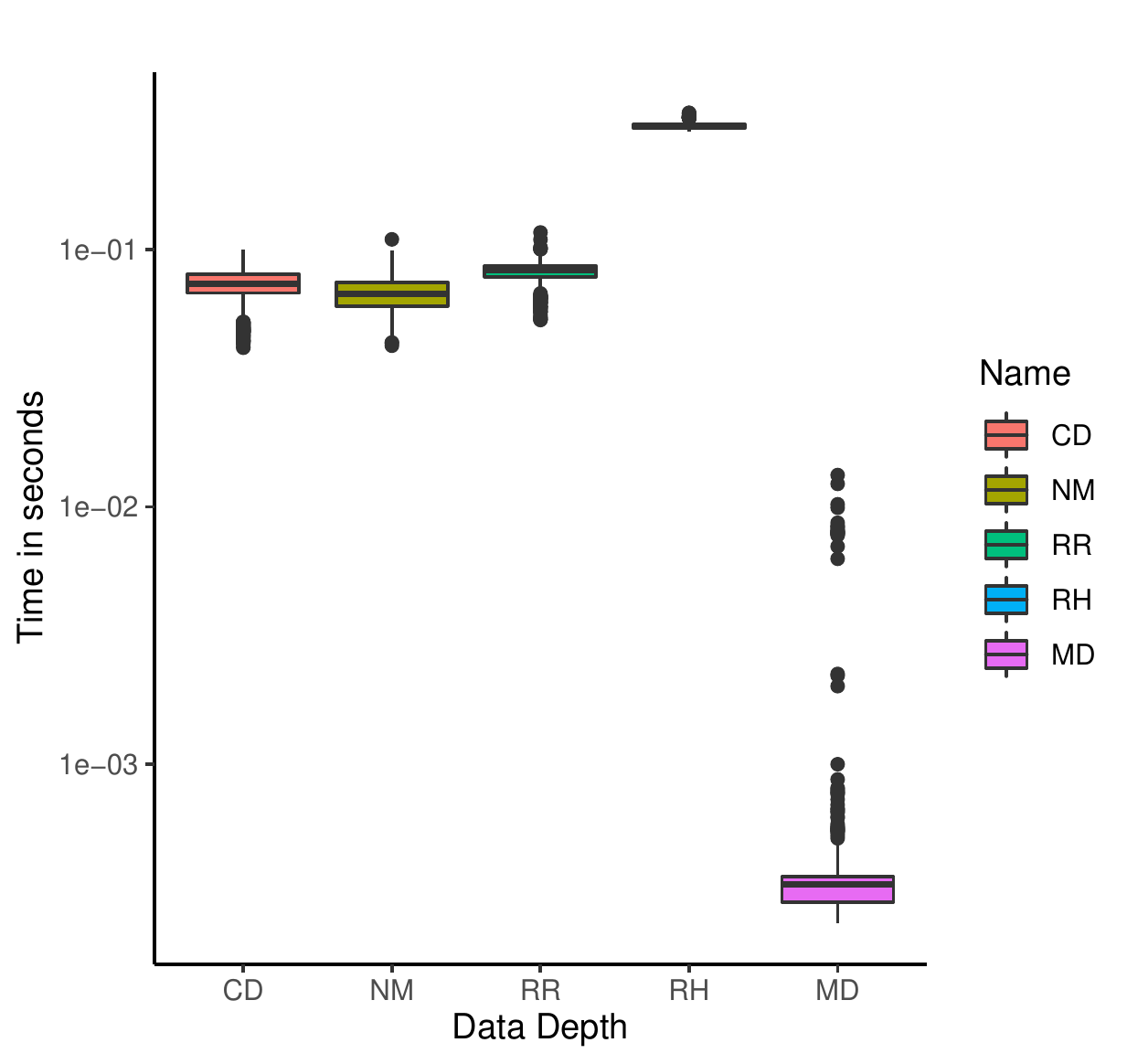} }}%
    \subfloat[\centering Merged reference samples of size $|R| = 1600$]{{\includegraphics[width=0.49\textwidth]{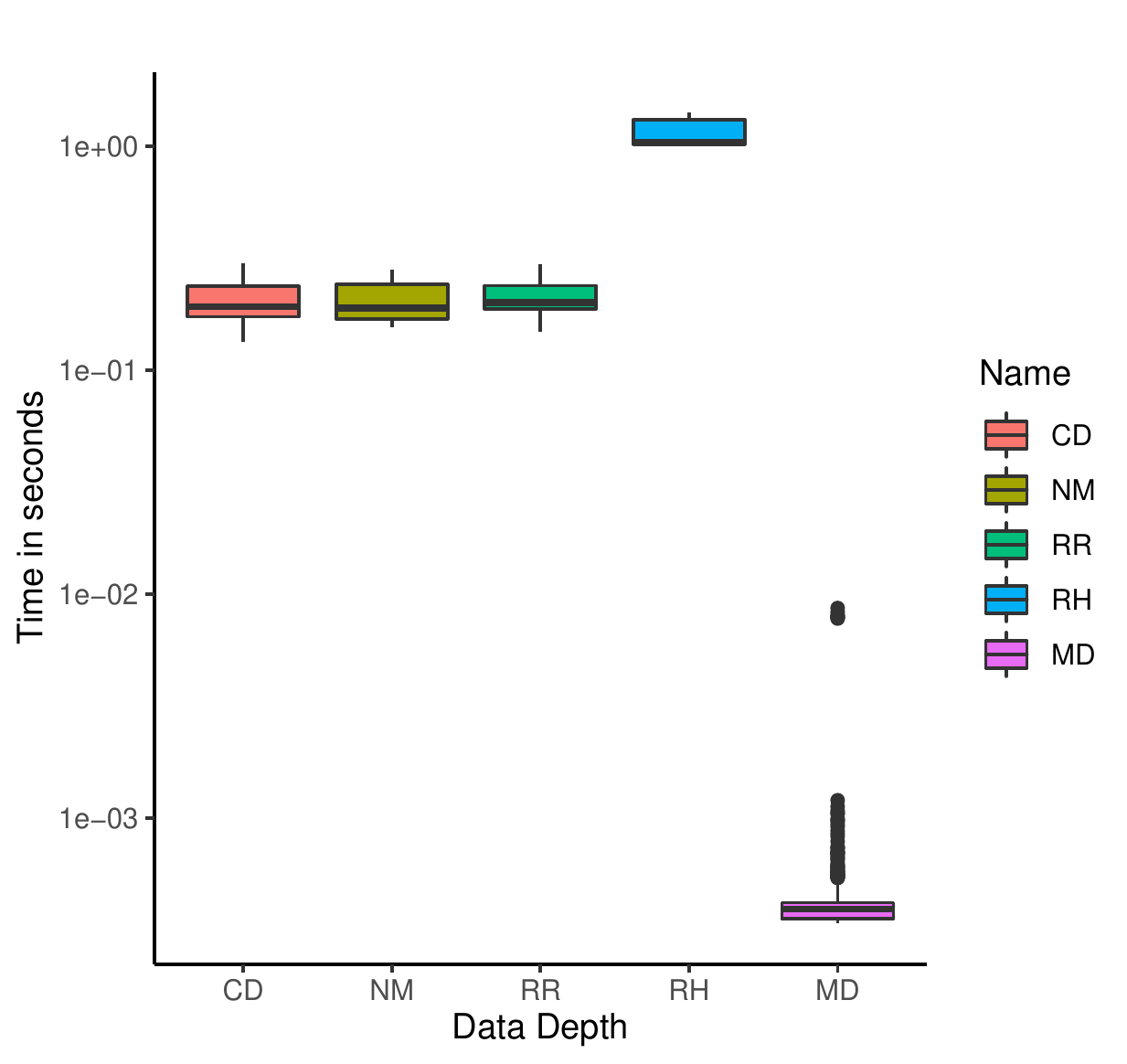} }}%
    \caption{Distribution of computation time for different data depths in Experiment 2 on a logarithmic scale. The order is symmetric Projection Depth with \textbf{C}oordinate \textbf{D}escent, \textbf{N}elder-\textbf{M}ead and \textbf{R}efined \textbf{R}andom  algorithms, \textbf{R}obust \textbf{H}alfspace and \textbf{M}ahalanobis \textbf{D}epths.}%
    \label{Exp2Time}%
\end{figure}

\begin{figure}%
    \centering
    \subfloat[\centering Single class reference samples of size $|R| = 60$]{{\includegraphics[width=0.49\textwidth]{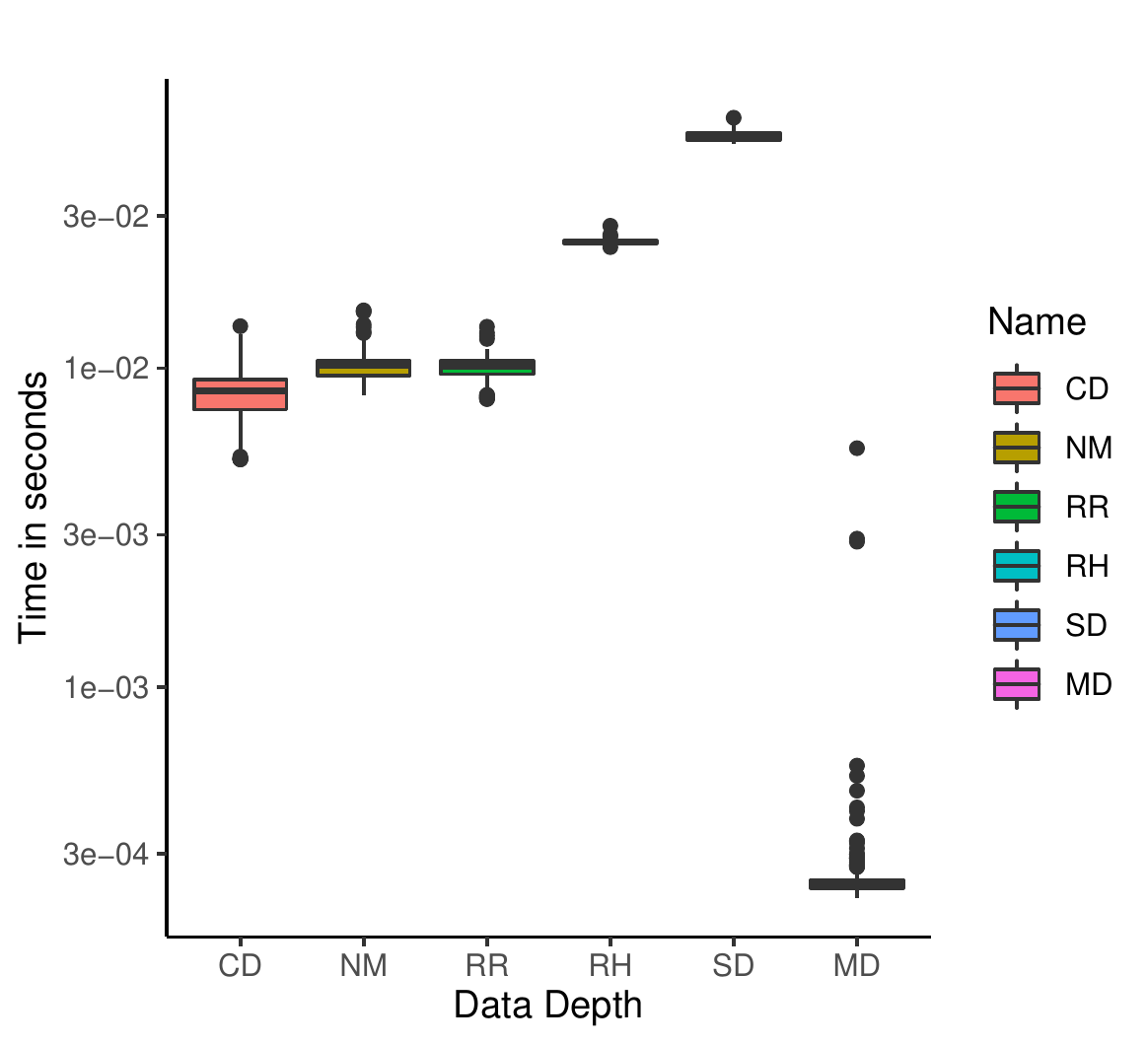} }}%
    \subfloat[\centering Merged reference samples of size $|R| = 140$]{{\includegraphics[width=0.49\textwidth]{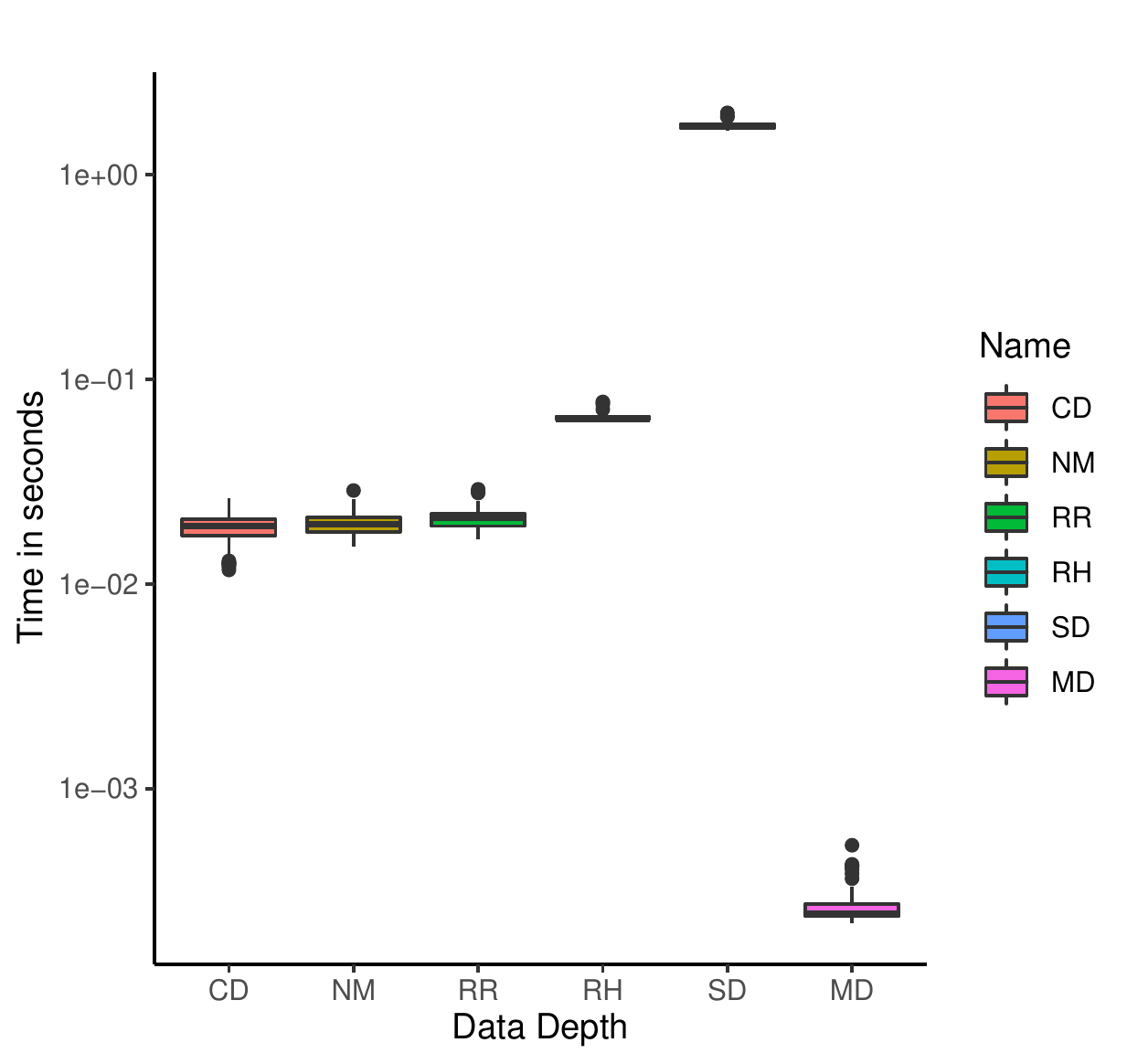} }}%
    \caption{Distribution of computation time for different data depths in Experiment 3 on a logarithmic scale. The order is symmetric Projection Depth with \textbf{C}oordinate \textbf{D}escent, \textbf{N}elder-\textbf{M}ead and \textbf{R}efined \textbf{R}andom  algorithms, \textbf{R}obust \textbf{H}alfspace, \textbf{S}implicial and \textbf{M}ahalanobis \textbf{D}epths.}%
    \label{Exp3Time}%
\end{figure}

In Figure \ref{Exp3Time}, we can see that Simplicial depth requires considerably longer to be computed than other notions of data depth. Regarding the algorithms to approximate Projection depth, the running times remain similar in Experiments 1 and 2, Figures \ref{Exp1Time} and \ref{Exp2Time}, respectively. Despite the increased complexity of experiments, Mahalanobis depth is characterized by a stable and low running time. On the contrary, the running time of \textbf{HD}$_r$ increases noticeably with the growing size of reference samples as well as additional dimensions. 

To summarize, if we exclude the performance of \textbf{MD}, the computation of data depth in $\mathbb{R}^{16}$ for one point with $|R| = 3000$ would usually take more than 10 seconds. In statistics, such results seem to be acceptable. However, taking into account the current applications of ANNs, for example, the image classification applying a CNN, the time required to process one image is under 0.1 second  (cf. \citealp{shi2022remote}). Thus, we should critically consider the running times for the computation of data depth, striving for their improvements and guarantee the applicability of the proposed framework to monitor state-of-the-art models based on AI by improving the software for data depth computation.

\end{document}